\documentclass[preprint,authoryear,sort&compress,12pt]{elsarticle}  
\usepackage[top=1.25in, bottom=1.25in, left=1.0in, right=1.0in]{geometry}
\usepackage{graphicx}
\usepackage{amssymb}
\usepackage{amsmath}
\usepackage{lineno}
\usepackage{subfig}
\usepackage{array}

\usepackage{color}
\usepackage{multirow}

\newcommand{\xmb}[1]{\ensuremath{\mathbf{#1}}}
\newcommand{\xmbs}[1]{\ensuremath{\boldsymbol{#1}}}

\journal{Computers \& Geosciences}

\begin{document}

\begin{frontmatter}

 \title{SediFoam: A general-purpose, open-source CFD--DEM solver for particle-laden flow with
 emphasis on sediment transport}

 \author{Rui Sun} \ead{sunrui@vt.edu}
 \author{Heng Xiao\corref{corxh}} \ead{hengxiao@vt.edu}
 \address{Department of Aerospace and Ocean Engineering, Virginia Tech, Blacksburg, VA 24060, United
 States}

 \cortext[corxh]{Corresponding author. Tel: +1 540 315 6242}

\begin{abstract}

\begin{sloppypar}
  With the growth of available computational resource, CFD--DEM (computational fluid
  dynamics--discrete element method) becomes an increasingly promising and feasible approach for the
  study of sediment transport. Several existing CFD--DEM solvers are applied in chemical engineering
  and mining industry. However, a robust CFD--DEM solver for the simulation of sediment transport is
  still desirable. In this work, the development of a three-dimensional, massively parallel, and
  open-source CFD--DEM solver \textit{SediFoam} is detailed. This solver is built based on
  open-source solvers OpenFOAM and LAMMPS. OpenFOAM is a CFD toolbox that can perform
  three-dimensional fluid flow simulations on unstructured meshes; LAMMPS is a massively parallel
  DEM solver for molecular dynamics. Several validation tests of \textit{SediFoam} are performed
  using cases of a wide range of complexities. The results obtained in the present simulations are
  consistent with those in the literature, which demonstrates the capability of \textit{SediFoam}
  for sediment transport applications. In addition to the validation test, the parallel efficiency
  of \textit{SediFoam} is studied to test the performance of the code for large-scale and complex
  simulations. The parallel efficiency tests show that the scalability of \textit{SediFoam} is
  satisfactory in the simulations using up to $O(10^7)$ particles.
\end{sloppypar}

\end{abstract}


\end{frontmatter}


\section{Introduction}
\label{sec:intro}

Particle-laden flows are frequently encountered in engineering applications such as coastal sediment
transport, gas-solid fluidization, and aerosol deposition. Numerical simulations of these systems
can improve the physical understanding of these flows. With the development of available
computational resources, the CFD--DEM approach becomes an increasingly promising approach for
particle-laden flows. In CFD--DEM, DEM approach tracks the motion of Lagrangian particles based on
Newton's second law; CFD solves the motion of fluid flow based on locally-averaged Navier--Stokes
equations~\citep{anderson67}. 

The CFD--DEM solvers include commercial solvers, research codes, and open-source solvers. These
solvers solve similar equations for fluid flow and particle motion, and use similar submodels (i.e.,
drag force model and particle collision model). {\color{black} Commercial solvers, such as Fluent
EDEM and the CFD--DEM packages in STAR-CCM+ and
AVL-Fire~\citep{spogis08mm,fries2011cfd,ebrahimi14cfd,eppinger2011cfd,jajcevic2013large}, are
general-purposed solvers without emphasis on sediment transport.} In-house research
codes~\citep{calantoni04ms,deb13an,capecelatro13ae,wu14parallel} can also be applied to CFD--DEM
simulations, but the accessibility of these solvers is limited. On the other hand, open-source
solvers~\citep{garg12os,li12os,goniva09tf} provide the user with good versatility in the development
of the numerical model.

The efforts in the development of CFD--DEM solver focus on gas-solid flows. However, several unique
features in sediment transport should be accounted for. First, for subaqueous sediment transport,
the lubrication and the added mass force on the particle are much larger than gas-solid flows. The
influences of them should be accounted for in CFD--DEM simulations. Moreover, sediment transport
occurs at the boundary layer. To resolve the fluid flow, the size of CFD mesh can be smaller than
sediment particle. Therefore, a robust algorithm is required when converting the properties of
discrete particles to the Eulerian CFD mesh. Finally, the parallel efficiency of the code is
important since the number of sediment particles can be as large as $O(10^6)$ in the simulation of
laboratory-scale problems.  {\color{black} Because these unique features of sediment transport are
ignored in previous numerical
simulations~\citep{drake01dp,calantoni04ms,duran12ns,furbish13ap,schmeeckle14ns}, a robust
open-source solver is still desirable for the study of sediment transport. In this work, a
three-dimensional, massively parallel, and open-source CFD--DEM solver \textit{SediFoam} with
emphasis on sediment transport is presented. The originality of \textit{SediFoam} includes: (a) the
lubrication and the added mass force on the particle; (b) the averaging algorithm to map the
properties of Lagrangian particles to Eulerian mesh~\citep{sun15db1}; (c) the parallel algorithm and
the performance test.  The potential advantage of \textit{SediFoam} include the ability to simulate
polydispersed particle mixtures, and the flexibility of applying turbulence models because of the
robustness of OpenFOAM, among others~\citep{wang14les}. } The present solver is developed based on
two state-of-the-art open-source solvers, i.e., CFD solver OpenFOAM (Open Field Operation and
Manipulation), and molecular dynamics simulator LAMMPS (Large-scale Atomic/Molecular Massively
Parallel Simulator). The two solvers are selected because they are both open-source, parallelized,
highly modular and well established~\citep{openfoam,lammps}.

The rest of this paper is organized as follows. The methodology of the present code is introduced in
Section~\ref{sec:cfddem}, including the mathematical formulation of fluid equations, particle motion
equations, and fluid--particle interactions. Section~\ref{sec:implementation} describes the
implementations, including the communication and the parallelization of the code. The numerical
validations of the code for various sediment transport problems are performed in
Section~\ref{sec:simulations}. The parallel efficiency of the present solver is detailed in
Section~\ref{sec:parallel-eff}. Finally, Section~\ref{sec:conclude} concludes the paper.

\section{Methodology}
\label{sec:cfddem}

\subsection{Mathematical Model of Particle Motion}
\label{sec:dem-particle}

In the CFD--DEM approach, the translational and rotational motion of each particle is calculated
based on Newton's second law as the following equations~\citep{cundall79,ball97si}: 
\begin{subequations}
 \label{eq:newton}
 \begin{align}
  m \frac{d\xmb{u}}{dt} &
  = \xmb{f}^{col} + \xmb{f}^{lub} + \xmb{f}^{fp} + m \xmb{g} \label{eq:newton-v}, \\
  I \frac{d\xmbs{\Psi}}{dt} &
  = \xmb{T}^{col} + \xmb{T}^{lub} + \xmb{T}^{fp} \label{eq:newton-w},
 \end{align}
\end{subequations}
where \( \xmb{u} \) is the velocity of the particle; $t$ is time; $m$ is particle mass;
\(\xmb{f}^{col} \) represents the contact forces due to particle--particle or particle--wall
collisions; \(\xmb{f}^{lub} \) is the lubrication force due to the fluid squeezed out from the gaps
between two particles; \(\xmb{f}^{fp}\) denotes fluid--particle interaction forces (e.g., drag, lift
force, added mass force, and buoyancy); \(\xmb{g}\) denotes body force. Similarly, \(I\) and
\(\xmbs{\Psi}\) are angular moment of inertia and angular velocity, respectively, of the particle;
\(\xmb{T}^{col}\), \(\xmb{T}^{lub}\) and \(\xmb{T}^{fp}\) are the torques due to contact forces,
lubrication, and fluid--particle interactions, respectively.  To compute the collision forces and
torques, the particles are modeled as soft spheres with inter-particle contact represented by an
elastic spring and a viscous dashpot. Further details can be found in~\cite{tsuji93}.

\subsection{Locally-Averaged Navier--Stokes Equations for Fluids}
\label{sec:lans}

The fluid phase is described by the locally-averaged incompressible Navier--Stokes equations.
Assuming constant fluid density \(\rho_f\), the governing equations for the fluid
are~\citep{anderson67,kafui02}:
\begin{subequations}
 \label{eq:NS}
 \begin{align}
  \nabla \cdot \left(\varepsilon_s \xmb{U}_s + {\varepsilon_f \xmb{U}_f}\right) &
  = 0 , \label{eq:NS-cont} \\
  \frac{\partial \left(\varepsilon_f \xmb{U}_f \right)}{\partial t} + \nabla \cdot \left(\varepsilon_f \xmb{U}_f \xmb{U}_f\right) &
  = \frac{1}{\rho_f} \left( - \nabla p + \varepsilon_f \nabla \cdot \xmbs{\mathcal{R}} + \varepsilon_f \rho_f \xmb{g} + \xmb{F}^{fp}\right), \label{eq:NS-mom}
 \end{align}
\end{subequations}
where \(\varepsilon_s\) is the solid volume fraction; \( \varepsilon_f = 1 - \varepsilon_s \) is the
fluid volume fraction; \( \xmb{U}_f \) is the fluid velocity. The terms on the right hand side of
the momentum equation are: pressure (\(p\)) gradient, divergence of the stress tensor \(
\xmbs{\mathcal{R}} \) , gravity, and fluid--particle
interactions forces, respectively.  {\color{black} Large eddy simulation is performed in the present
work, the stress tensor is composed of both viscous and Reynolds stresses \( \xmbs{\mathcal{R}} =
\mu\nabla \xmbs{U_f} + \rho_f\xmbs{\tau}\), where $\mu$ is the dynamic viscosity of the fluid flow;
$\xmbs{\tau}$ is the Reynolds stress. The expression of the Reynolds stress is:
\begin{equation}
  \xmbs{\tau} = \frac{2}{\rho_f}\mu_t \xmbs{S} - \frac{2}{3}k\xmbs{I};
  \label{eq:reynolds-stress}
\end{equation}
where $\mu_t$ is the dynamics eddy viscosity; $\xmbs{S} = (\nabla \xmbs{U_f} + (\nabla
\xmbs{U_f})^T)/2$; $k$ is the turbulent kinetic energy.  It is noted that in the stress tensor \(
\xmbs{\mathcal{R}} \) term, the fluctuations of the fluid flow at the boundary of the particle are
not resolved. 
}
The Eulerian fields $\varepsilon_s$, $\xmb{U}_s$, and
$\xmb{F}^{fp}$ in Eq.~(\ref{eq:NS}) are obtained by averaging the information of Lagrangian
particles. {\color{black} However, the fluid variables are not averaged before sending to the
particle, which is because the averaged fluid variables might be too diffusive for calculation.}
{\color{black} It is noted that the coupling of the fluid and solid phases is similar with Scheme 3
according to the literature~\cite{feng04am,zhou10dps}. At each time step, the fluid-particle
interaction forces on individual particles are calculated first in the CFD cell, and the values are
then summed to produce the interaction force at the cell scale. Detailed information of the coupling
between the solvers can be found in Section~\ref{sec:implementation}.  }

\subsection{Fluid--Particle Interactions}
\label{sec:fpi}
The fluid-particle interaction force \(\xmb{F}^{fp}\) consists of buoyancy \( \xmb{F}^{buoy} \),
drag \( \xmb{F}^{drag} \), lift force \(\xmb{F}^{lift}\), and added mass force \(\xmb{F}^{add}\).
Although the lift force and the added mass force are usually ignored in fluidized bed simulation,
they are important in the simulation of sediment transport. 

The drag on an individual particle $i$ is formulated as:
\begin{equation}
  \mathbf{f}^{drag}_i = \frac{V_{p,i}}{\varepsilon_{f, i} \varepsilon_{s, i}} \beta_i \left( \mathbf{u}_{p,i} -
  \mathbf{U}_{f, i} \right),
  \label{eqn:particleDrag}
\end{equation}
where \( V_{p, i} \) and \( \mathbf{u}_{p, i} \) are the volume and the velocity of particle $i$,
respectively; \( \mathbf{U}_{f, i} \) is the fluid velocity interpolated to the center of particle
$i$; \( \beta_{i} \) is the drag correlation coefficient which accounts for the presence of other
particles.
{\color{black}
The $\beta_i$ value for the drag force is based on~\cite{mfix93}:
\begin{equation}
\beta_i = \frac{3}{4}\frac{C_d}{V^2_r}\frac{\rho_f |\xmbs{u_{p,i}} - \xmbs{U_{f,i}}|}{d_{p,i}}
\mathrm{, \quad with \quad} C_d = \left( 0.63+0.48\sqrt{V_r/\mathrm{Re}} \right),
  \label{eqn:beta-i}
\end{equation}
where the particle Reynolds number Re is defined as:
\begin{equation}
  \mathrm{Re} = \rho_i d_{p,i} |\xmbs{u_{p,i}} - \xmbs{U_{f,i}}|;
  \label{eqn:p-re}
\end{equation}
the $V_r$ is the correlation term:
\begin{equation}
  V_r = 0.5\left( A_1 - 0.06\mathrm{Re}+\sqrt{(0.06\mathrm{Re})^2 +
  0.12\mathrm{Re}(2A_2 - A_1)+A_1^2} \right),
  \label{eqn:drag-vr}
\end{equation}
with
\begin{equation}
  A_1 = \varepsilon_f^{4.14}, \quad
  A_2 =
  \begin{cases}
    0.8\varepsilon_f^{1.28} & \quad \text{if } \varepsilon_f \le 0.85, \\
    \varepsilon_f^{2.65}    & \quad \text{if } \varepsilon_f > 0.85.\\
  \end{cases}
  \label{eqn:drag-A}
\end{equation}
Note that other drag force models are also implemented in~\textit{SediFoam} with correlations for
$\beta$ in dense particle-laden flows~\citep{di94vo,wen13me}.} In addition to drag, the lift force
on a spherical particle is modeled in \textit{SediFoam} as~\citep{saffman65th,rijn84se1}:
\begin{equation}
  \mathbf{f}_{i}^{lift} = C_{l} \rho_f \nu^{0.5} D^{2} \left( \mathbf{u}_{p,i} - \mathbf{U}_{f,i}
  \right) \boldsymbol{\times} \nabla \mathbf{U}_{f,i},
  \label{eqn:particleLift}
  \end{equation}
where $\boldsymbol{\times}$ indicates the cross product of two vectors, and $C_{l} = 1.6$ is the
lift coefficient. The added mass force is considered in \textit{SediFoam} due to the comparable
densities of the carrier and disperse phases in sediment transport applications. This is modeled as:
\begin{equation}
  \mathbf{f}_{i}^{add} = C_{add} \rho_f V_{p,i} \left( \frac{\mathrm{D}\mathbf{u}_{p,i}}{\mathrm{D}t}
  - \frac{\mathrm{D}\mathbf{U}_{f,i}}{\mathrm{D}t} \right),
  \label{eqn:particleAddedMass}
  \end{equation}
where $C_{add} = 0.5$ is the coefficient of added mass.  {\color{black} The Lagrangian particle
acceleration term ${\mathrm{D}\mathbf{u}_{p,i}}/{\mathrm{D}t}$ utilizes the particle velocity at the
previous time; the material derivative of fluid phase velocity
${\mathrm{D}\mathbf{U}_{f,i}}/{\mathrm{D}t} = {\mathrm{d}\mathbf{U}_{f,i}}/{\mathrm{d}t} +
\mathbf{U}_{f,i}\cdot\nabla\mathbf{U}_{f,i}$ can be obtained in OpenFOAM at each time step.  The
lift force and added mass models used in the present work is only applicable for particle-laden
flows in small Reynolds number and low volume fraction regime. The study of~\cite{rijn84se1} has
indicated the accuracy of such modeling scheme is acceptable for sediment transport applications.  }

\subsection{Diffusion-Based Averaging Method}

\label{sec:cg}
According to Eq.~(\ref{eq:NS}), the continuum Eulerian fields of \( \varepsilon_s \), \( \xmb{U}_s
\), and \( \xmb{F}^{fp} \) are obtained by averaging from discrete particle data. In the present
solver, the averaging algorithm previously proposed by the authors~\citep{sun15db1,sun15db2} is
implemented.

Taking the averaging process of $\varepsilon_s$ as an example. In the first step, the particle
volumes at each CFD cell are obtained. Then, the solid volume fraction for cell \(k\) is calculated by
dividing the total particle volume by the volume of this cell \(V_{c, k}\). That is:
\begin{equation}
  \varepsilon_{s,k}(\mathbf{x},\tau = 0)
 = \frac{\sum_{i=1}^{n_{p, k}} V_{p, i}}{V_{c, k}} , \label{eq:pcm-k} 
\end{equation}
where \(n_{p, k}\) is the number of particles in cell \(k\). With the initial condition in
Eq.~(\ref{eq:pcm-k}), a transient diffusion equation for $\varepsilon_s(\mathbf{x}, \tau)$ is solved
to obtain the continuum Eulerian field of $\varepsilon_s$: \begin{equation} \frac{\partial
  \varepsilon_s}{\partial \tau} = \nabla^2 \varepsilon_s
\label{eq:diffusion-c}
\end{equation}
where \(\nabla^2 \) is the Laplacian operator; \(\tau\) is pseudo-time. It has been established
in~\citet{sun15db1, sun15db2} that the results obtained by Eq.~(\ref{eq:diffusion-c}) is
equivalent with Gaussian kernel based averaging with bandwidth $b = \sqrt{4\tau}$. Similarly, the
smoothed \(\xmb{U}_s\) and \(\xmb{F}^{fp}\) fields can be obtained by using this approach. 

{\color{black} 
We note that a similar averaging procedure has been proposed earlier by~\cite{capecelatro13ae},
where a mollification procedure with Gaussian kernel is followed by solving a diffusion equation of
the obtained volume fraction. Both methods are conservative and mesh-independent, and are
theoretically equivalent. The novelty of~\cite{sun15db1, sun15db2} lies in the following aspects.
First, they established the theoretical equivalence between the diffusion based averaging procedure
and the Gaussian kernel averaging commonly used in statistical mechanics~\citep{zhu02ave}. Based on
this insight, they provided a clear physical interpretation of  the normalized diffusion time,
rendering it a physical parameter related to the wake of the particles.  Second, the wall boundaries
are treated in~\cite{sun15db1} with a straightforward, efficient no-flux boundary condition, and the
conservativeness has been proved analytically.  In contrast,~\cite{capecelatro13ae} ensured the
conservativeness of the averaging procedure at wall boundaries by using ghost particles. Finally,
the coarse-graining procedure of~\cite{sun15db1, sun15db2} is implemented based on the open-source,
general-purpose, three-dimensional, massively parallel CFD solver with a generic unstructured
body-fitting mesh, while~\cite{capecelatro13ae} used an in-house CFD solver based on a Cartesian
mesh with an immersed boundary method. More details of the averaging algorithm can be found
in~\cite{sun15db1, sun15db2}.
}

\subsection{Numerical Methods}
\label{sec:num-method}
The solution of the particle motions including their interactions via collisions and endured
contacts are handled by LAMMPS. The fluid forces \( \xmb{f}^{fp} \) on the particles are computed in
OpenFOAM, supplied into LAMMPS, and used in the integration of particle motion
equations~(\ref{eq:newton}). The particle forces in the fluid equations are computed in OpenFOAM
according to the forces on individual particles via the averaging procedure.

The fluid equations in~(\ref{eq:NS}) are solved by OpenFOAM using the finite volume method
\citep{jasak96ea}. The discretization is based on a collocated grid, i.e., pressure and all velocity
components are stored in cell centers. PISO (Pressure Implicit Splitting Operation) algorithm is
used for velocity--pressure decoupling~\citep{issa86so}. Second-order central schemes are used for
the spatial discretization of convection terms and diffusion terms. Time integrations are performed
using a second-order implicit scheme. The averaging method involves solving transient diffusion
equations based on the OpenFOAM platform. The diffusion equations are also solved on the CFD mesh. A
second-order central scheme is used for the spatial discretization of the diffusion equation; a
second-order implicit scheme is used for the temporal integration.

{\color{black}
To solve the equation of motion of the particles in (\ref{eq:newton}), the nearest particles of each
particle are tracked. To find the nearest particles, LAMMPS uses a combination of neighbor lists and
link-cell binning~\citep{hockney74qh} and the scale of the computation is only
$O(N)$~\citep{lammps}.  The collision force is computed with a linear spring-dashpot model, in which
the normal elastic contact force between two particles is linearly proportional to the overlapping
distance~\citep{cundall79}. The lubrication model is based on~\cite{ball97si}, in which the force is
proportional to the relative velocity and inversely proportional to the relative distance.  The
torque on the particles due to the lubrication and collision are integrated to calculate the
particle rotation, but the interaction of the particle rotation and fluid is not considered.
Finally, the time step to resolve the particle collision is 1/50 the contact time to avoid particle
inter-penetration~\citep{sun07ht}.
}

\section{Implementations}
\label{sec:implementation}

\textit{SediFoam} was originally developed by the second author and his co-workers to study particle
segregation dynamics~\citep{sun09}. To improved the solver, we enhanced its parallel computing
capabilities and implemented the averaging algorithm for sediment transport. The source code of
\textit{SediFoam} is available at https://github.com/iurnus/SediFoam.git.

\begin{figure}[!htpb]
  \centering
  \includegraphics[natheight = 500, natwidth = 400,width=0.50\textwidth]{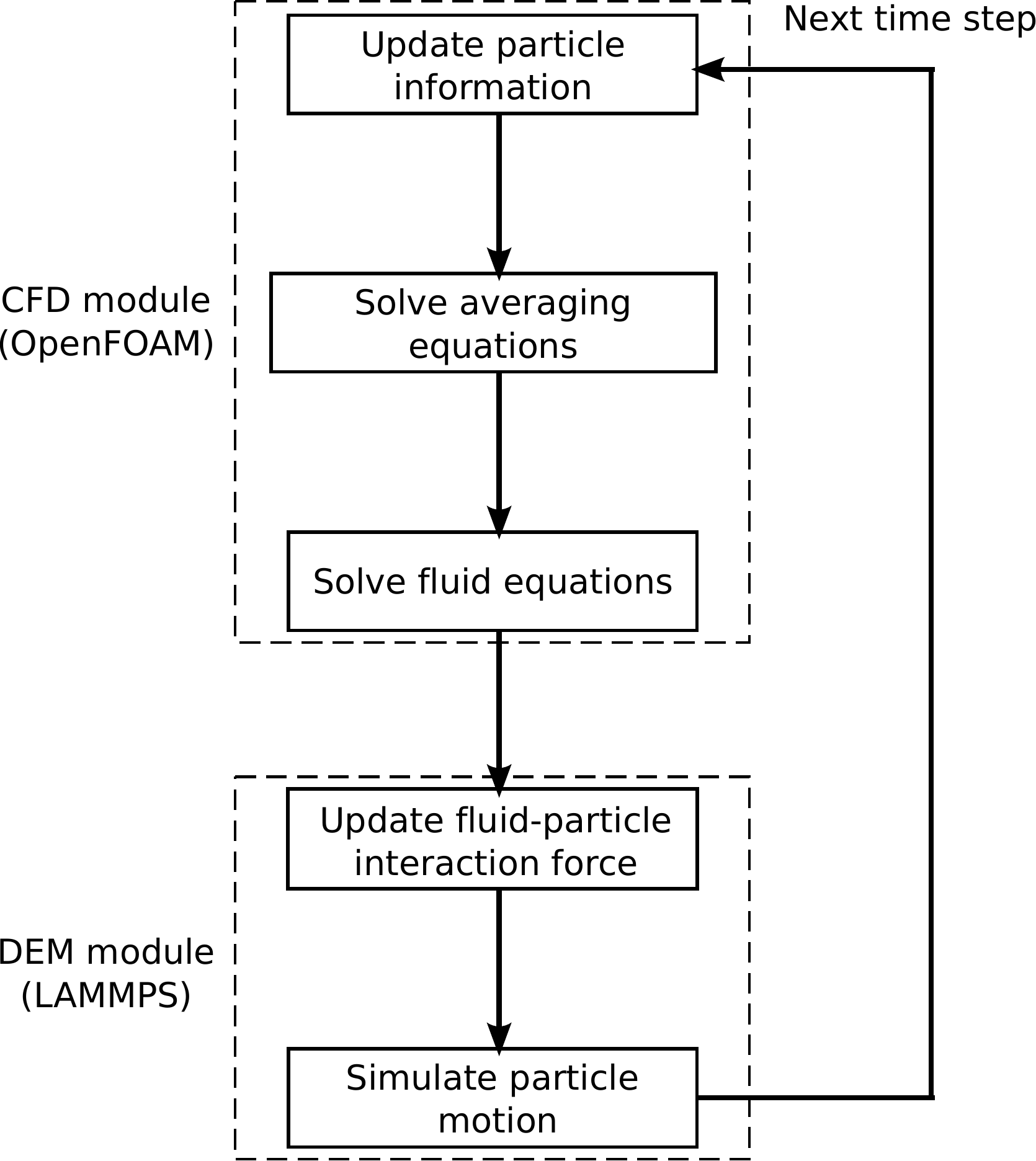}
  \caption{The block diagram of the code.}
  \label{fig:block-diagram}
\end{figure}

The diagram of the code is shown in Fig.~\ref{fig:block-diagram}. It can be seen in the figure that
the fluid and particle equations are solved individually by the CFD and DEM module at each time
step. The averaging procedure is performed in the CFD module before solving the fluid equations. In
addition to solving the equations, the information of the sediment particles is updated before CFD
module starts the averaging procedure; the fluid-particle interaction force of each particle is
updated the before DEM module evolves the motion of the particles. These procedures before solving
the fluid and particle equations are the coupling between CFD and DEM modules. 

\begin{figure}[!htpb]
  \centering
  \includegraphics[natheight = 500, natwidth = 800,width=0.9\textwidth]{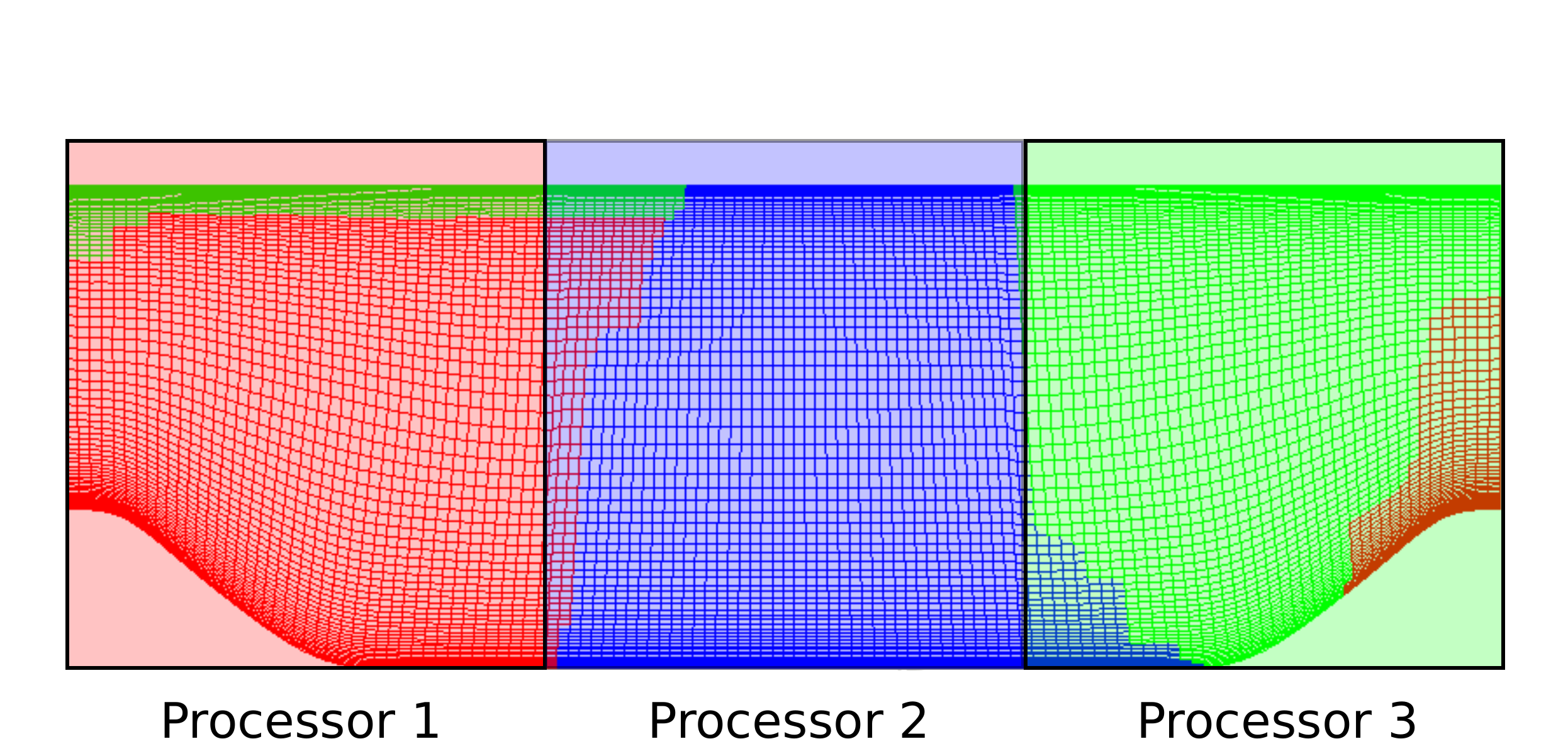}
  \caption{Decomposition of the computational domain in CFD--DEM simulation. Different colors in the
  CFD mesh denotes different subdomains in the CFD module, while the bound boxes of different colors
  denotes different subdomains in the DEM module.}
  \label{fig:domain-overlap}
\end{figure}

In the parallelization of \textit{SediFoam}, which is essential for simulating large granular
systems, the equations are solved in parallel by the DEM and CFD modules. Therefore, the
parallelization of \textit{SediFoam} concerns with the coupling of the modules between multiple
processors. When using multiple processors to accelerate the simulation, both modules decompose the
computational domain into $N_{proc}$ subdomains. In the coupling procedure, the particle information
obtained in each module is transfered to the other module. If the subdomains of individual
processors in each module were perfectly consistent, the information of every particle would be
local to each processor for both CFD and DEM modules. In this situation, inter-processor
communication is unnecessary in the coupling. However, the subdomains in most numerical simulations
are inconsistent (see Fig.~\ref{fig:domain-overlap}). This is due to the consideration of parallel
efficiency when decomposing the domain. In this situation, the information of some particles
obtained in CFD and DEM module are not stored in the same processor. Hence, inter-processor
communication is required for these non-local particles.

\begin{figure}[!htpb]
  \centering
  \includegraphics[natheight = 1200, natwidth = 1200,width=0.6\textwidth]{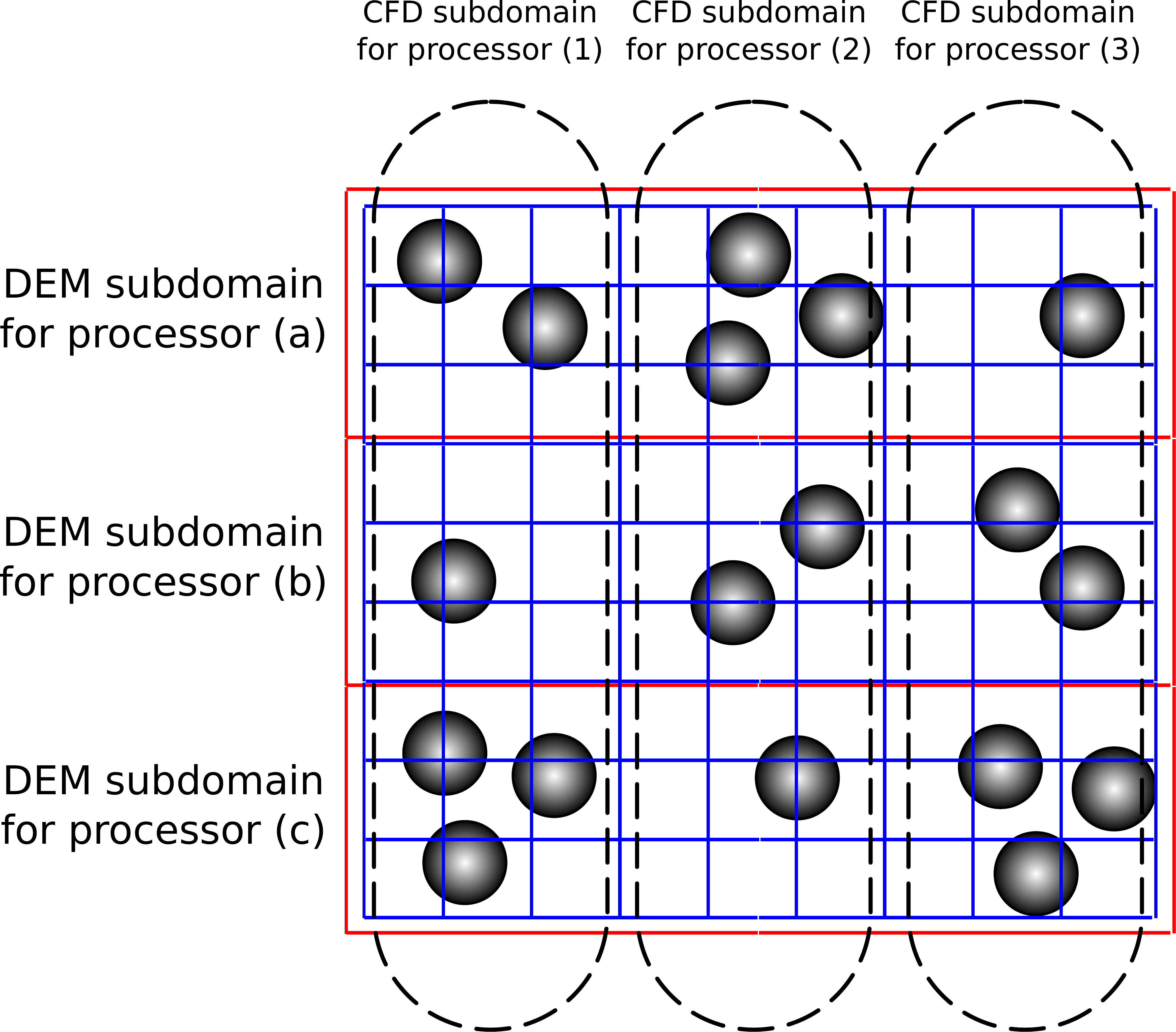}
  \caption{The geometry of a representative CFD--DEM case that the decompositions of the CFD and the
    DEM modules are different. The blue lines are the CFD mesh and the black dash lines illustrate
    the subdomains of the CFD module; the red lines are the subdomains of the DEM module.}
  \label{fig:domain-transpose-layout}
\end{figure}

\begin{figure}[!htpb]
  \centering
  \includegraphics[natheight = 2500, natwidth = 2500,width=0.9\textwidth]{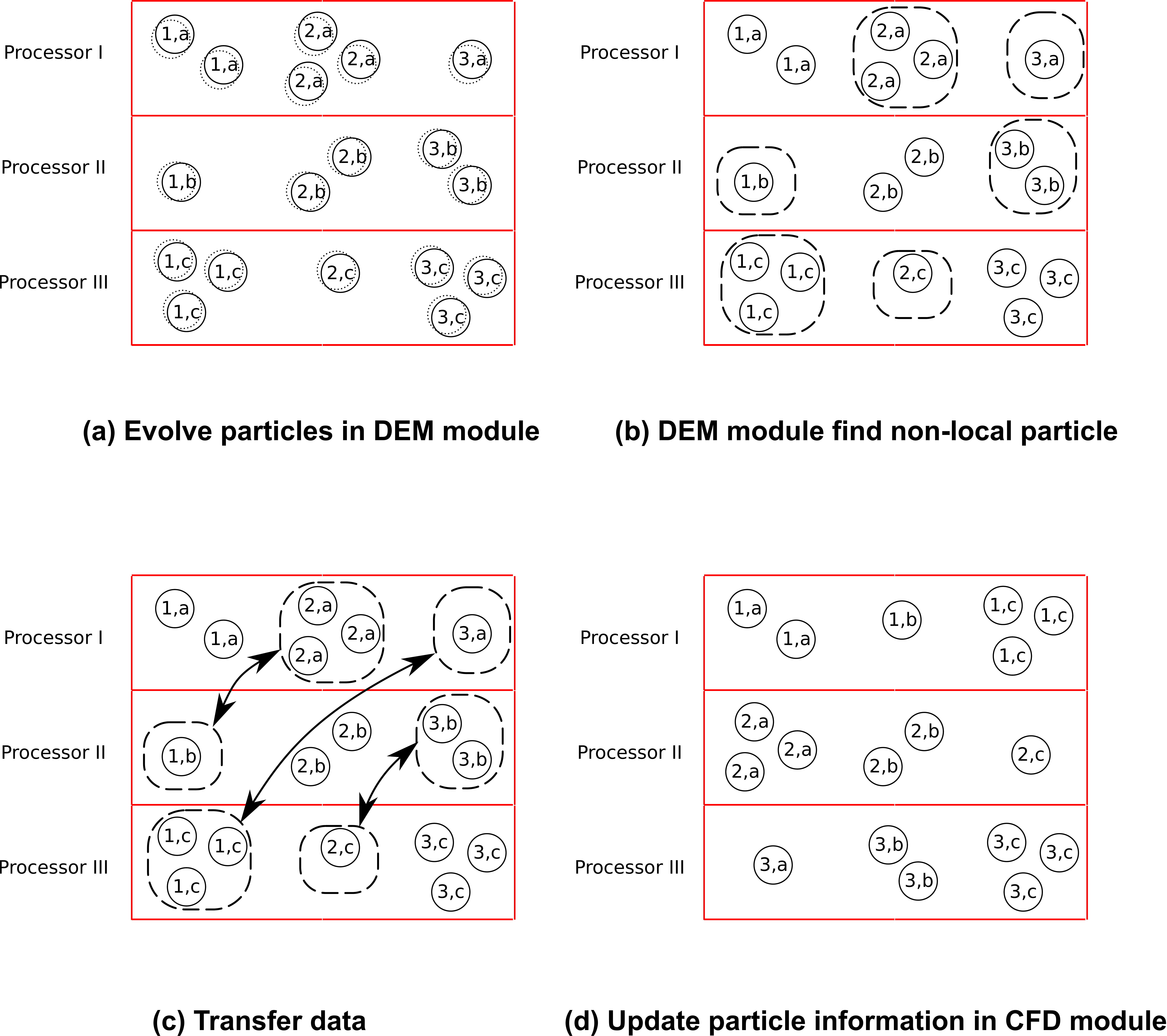}
  \caption{Inter-processor communication of the particle information. Each circle
    represents a sediment particle in the simulation. Number 1, 2, and 3 indicates the processor that
    stores the data of the particle in the CFD module; character ``a'', ``b'' and ``c'' denote the
    processor that store the data of the particle in the DEM module.}
  \label{fig:domain-transpose}
\end{figure}

An example is used to describe the parallelization of the coupling procedure in \textit{SediFoam}.
The geometry of this example employing three processors is shown in
Fig.~\ref{fig:domain-transpose-layout}. The parallelization in the present solver is performed as
follows:
\begin{enumerate}
\item
The DEM module of \textit{SediFoam} evolved the particles one step forward, shown in
Fig.~\ref{fig:domain-transpose}(a).
\item
The non-local particles are found in the DEM module, shown in Fig.~\ref{fig:domain-transpose}(b).
This is the preparation step before transferring non-local data.
\item
In each processor, the information of non-local particles is transfered to other
processors. This step is illustrated in Fig.~\ref{fig:domain-transpose}(c). 
\item
The particle information obtained in the DEM module is now local to the CFD module, shown in
Fig.~\ref{fig:domain-transpose}(d). The information of the particles in the CFD module is updated
and can be used in the next CFD step.
\end{enumerate}
After Step 4, the inter-processor communication is finished. The non-local data obtained by the DEM
module is transferred to update the information of sediment particles in the CFD module. Following
this approach, the non-local information obtained in the CFD module can also be transferred to the
particles in the DEM module.

\section{Results}
\label{sec:simulations}
Extensive validation tests of \textit{SediFoam} have been performed previously for fluidized bed
simulations~\citep{sun09,xiao-cicp,gupta15vav,sun15db2}. In the present work, three numerical tests
are presented to demonstrate the capability of \textit{SediFoam} in the simulation of sediment
transport. The sedimentation of a single particle in water is detailed in
Section~\ref{sec:pw-bounce}, which aims to validate the implementation of lubrication and added
mass. The motion of 500 particles on fixed sediment bed is discussed in
Section~\ref{sec:moderate-sediment}. The purpose of this case is to validate the properties of the
fluid and sediment particles obtained by using \textit{SediFoam}.  Simulations of relatively large
number of particles ($O(10^5)$) are detailed in Section~\ref{sec:large-sediment}. The objective of
this test is to demonstrate the capability of the present solver in the simulation of large and
complex cases.

\subsection{Case 1: Single Particle Sedimentation in Water}
\label{sec:pw-bounce}
Most numerical simulations using CFD--DEM are performed to study gas--solid flows. However, the
behavior of a sediment particle in liquid is different with that in gaseous flow. This is because
liquid has higher density and dynamic viscosity than gas. As such, lubrication and virtual mass
force of the same particle in liquid--solid flows are approximately $O(10^3)$ times larger than
those in gas--solid flows, and can be comparable to the weight of the sediment particle.  Therefore,
the influences of lubrication and virtual mass (usually negligible in gas--solid flows) can be
critical in the simulation of subaqueous sediment transport. To test the implementation of the two
forces in \textit{SediFoam}, a series of simulations of particle--wall collision are performed based
on the experiments of~\cite{gondret02bm}.

\begin{figure}[!htbp]
  \centering
    \subfloat[geometry of the domain]{
      \label{fig:layout-lub1}
      \includegraphics[natheight = 500, natwidth = 400,width=0.4\textwidth]{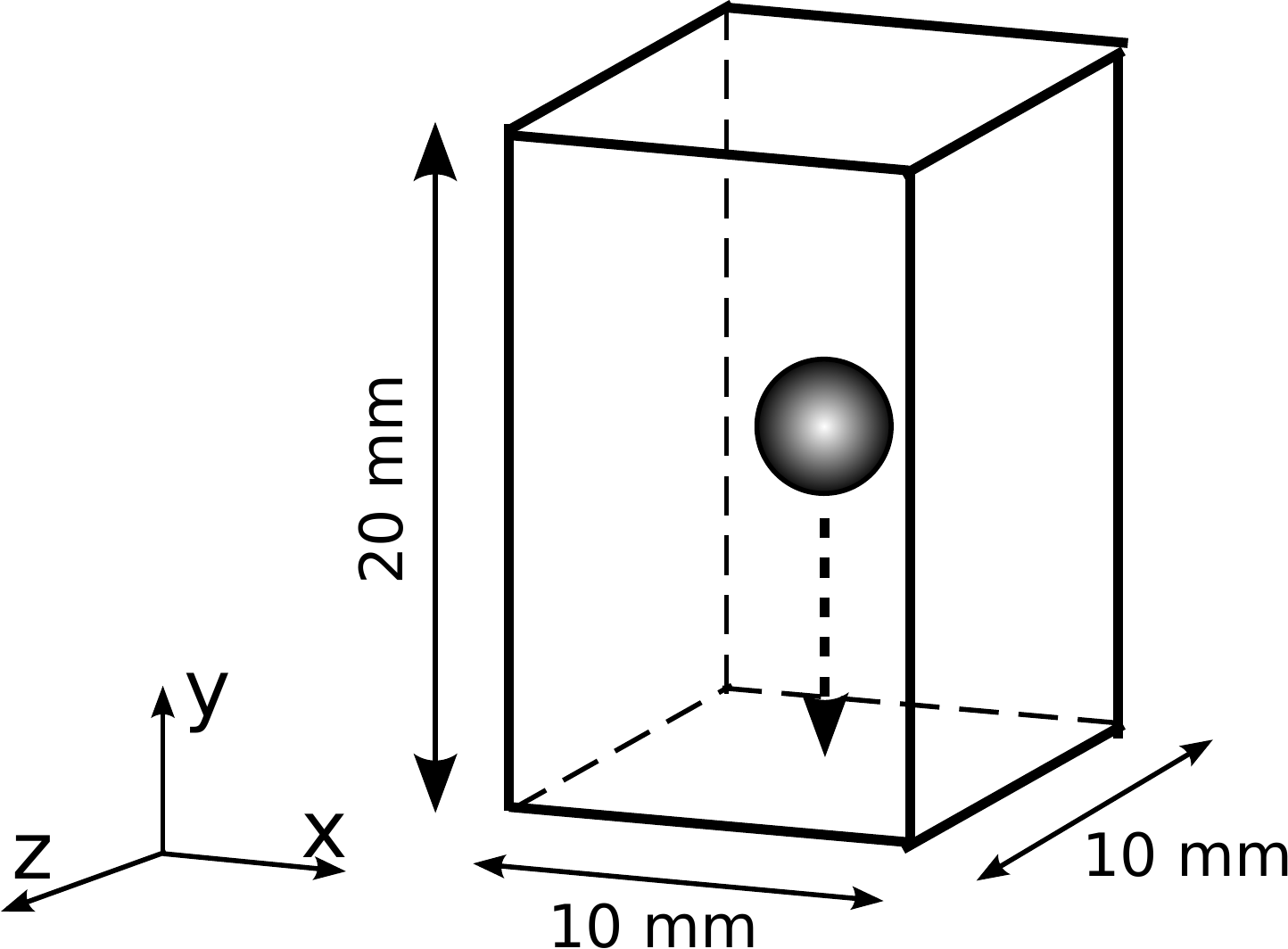}
    }
    \hspace{0.001\textwidth}
    \subfloat[snapshots of the particle positions]{
      \label{fig:layout-lub2}
      \includegraphics[natheight = 500, natwidth = 1600,width=0.9\textwidth]{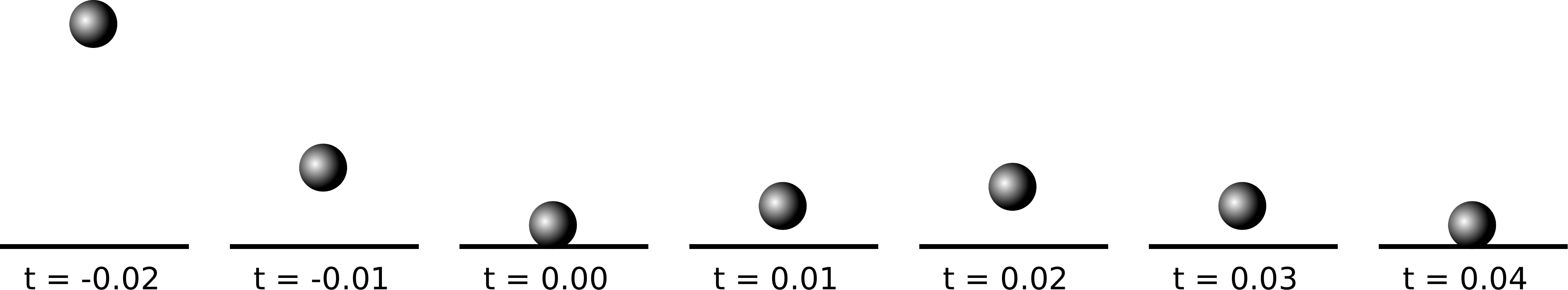}
    }
    \caption{(a) Geometry of the domain and (b) snapshots of the positions of the particle in the
    collision test. The dash line in panel (a) illustrates the direction of particle motion. In
    panel (b), $t = 0$~s responds to the moment the particle hits the wall for the first time.}
  \label{fig:layout-lub}
\end{figure}

The geometry of the domain in the particle--wall collision test is shown in
Figure~\ref{fig:layout-lub} along with the coordinate system. The parameters used in this case are
detailed in Table~\ref{tab:param-all}. Periodic boundary condition is applied at the boundaries in
both $x$- and $z$-directions. In the simulation, the fluid is quiescent
initially and the particle falls in the vertical direction due to the gravity force.  The initial
particle--wall distance is large enough so that the particle accelerates to the terminal velocity
before the collision occurs.

\begin{figure}[!htbp]
  \centering
    \subfloat[$St = 27$]{
      \label{fig:lub-collision1}
      \includegraphics[natheight = 500, natwidth = 600,width=0.45\textwidth]{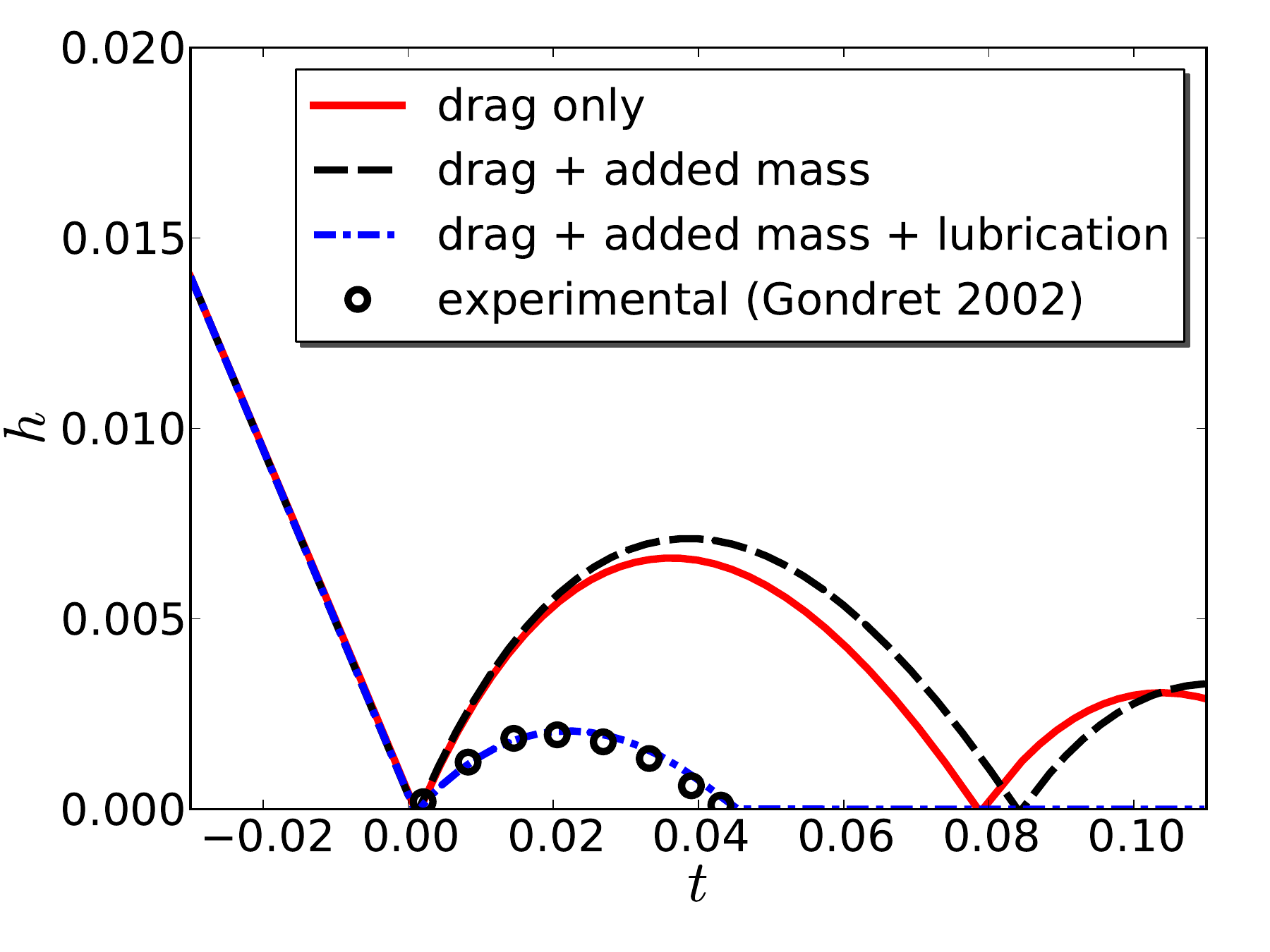}
    }
    \subfloat[$St = 742$]{
      \label{fig:lub-collision2}
      \includegraphics[natheight = 500, natwidth = 600,width=0.45\textwidth]{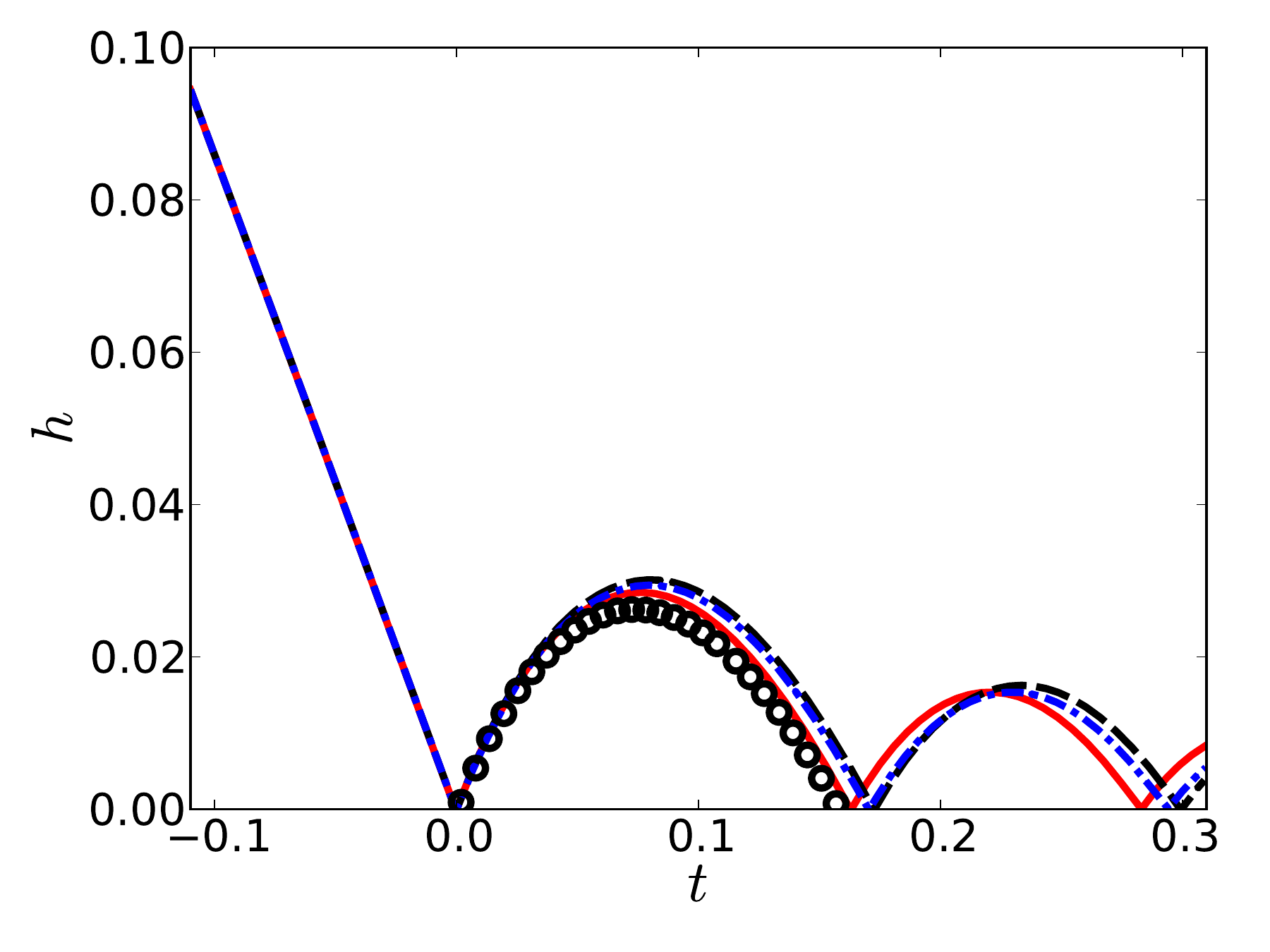}
    }
    \caption{The positions of the particle plotted as a function of time in the particle--wall
      bouncing test. The influence of lubrication and added mass are considered at Stokes number (a)
      $St = 27$ and (b) $St = 742$. \label{fig:lub-collision}}
\end{figure}

To test the influence of lubrication and added mass, the locations of the particle obtained in the
collision test are compared with the experimental results. The comparison is performed at two
different Stokes numbers, which is defined as:
\begin{equation}
  St = \frac{\rho_p D_p u_p}{9 \nu_f \rho_f} = \frac{\rho_p}{9\rho_f}Re_p.
  \label{eqn:stokes}
\end{equation}
In Fig.~\ref{fig:lub-collision}, the locations of the center of the sphere are plotted as a function
of time. It can be seen that the results obtained in the simulation that considered both lubrication
and added mass are consistent with the experimental measurements. Theoretically, added mass force
adds to the inertia of the particle, which leads to larger rebound height regardless of the Stokes
number. On the other hand, the lubrication depends on the viscous effect and decreases with Stokes
number. At $St = 27$, the viscous effect is large by definition. Consequently, the locations of the
particle predicted without lubrication are significantly different from the experimental data. At
$St = 742$, the viscous effects is small, so accounting for lubrication at large Stokes number does
not significantly influence the predictions of the particle motions.

\begin{figure}[!htbp]
  \centering
    \includegraphics[natheight = 500, natwidth = 800,width=0.6\textwidth]{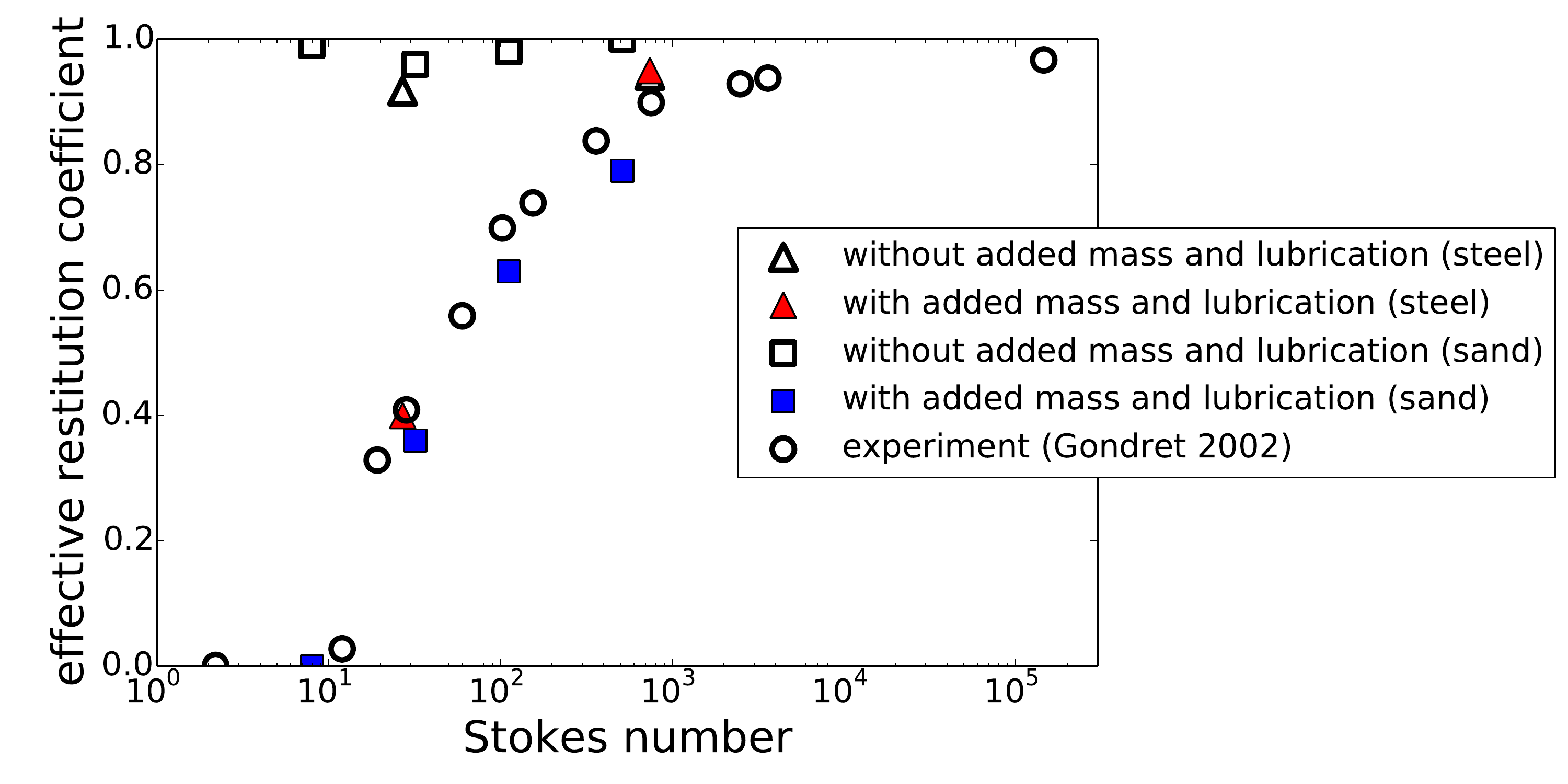}
    \caption{The influence of lubrication and added mass on the effective restitution coefficient in
    particle--wall collisions for different particles at different Stokes numbers.}
    \label{fig:lub-restitution}
\end{figure}

The prediction of the effective restitution coefficient using \textit{SediFoam} in particle--wall
bouncing test is shown in Fig.~\ref{fig:lub-restitution}. The effective restitution coefficient is
defined as $e = e_{water}/e_{air}$, where $e_{water}$ and $e_{air}$ are the restitution coefficients
for the same collision occurs in water and air, respectively. The amount of decrease of $e$ from 1
denotes the influence of lubrication and added mass in the collision. It can be seen in
Fig.~\ref{fig:lub-restitution} that the predictions from \textit{SediFoam} agree favorably with the
experimental data when the influence of lubrication and added mass is considered. In contrast, the
predictions without accounting for these forces deviate significantly from experimental measurements
by over-predicting the effective restitution coefficients, particularly at small Stokes number.
Accordingly, if sediment transport occurs at relatively low Stokes number~\citep{kidanemariam14id},
the influence of lubrication and added mass should be considered in CFD--DEM modeling of sediment
transport. Note that the calculation of lubrication incurs significant increase in computational
costs. Consequently, lubrication are accounted for by using a smaller dry restitution coefficient in
the literature~\citep{schmeeckle14ns,kidanemariam14dn}.

In summary, the influence of added mass and lubrication is important in the CFD--DEM modeling of
sediment transport. Additionally, the test cases validate the implementation of the added mass force
and lubrication force in \textit{SediFoam}.

\subsection{Case 2: Sediment Transport with 500 Particles}
\label{sec:moderate-sediment}
In this case, sediment transport at the boundary layer in the channel is studied. The results
obtained by using \textit{SediFoam} are compared with the numerical benchmark to illustrate the
capability of the present solver. The averaging algorithm proposed by~\cite{sun15db1} is applied to
obtain the continuum Eulerian quantities from discrete particles. This algorithm enables CFD--DEM
simulation at the boundary layer where the size of CFD cells is smaller than particles.

\begin{figure}[htbp]
  \centering
  \includegraphics[natheight = 500, natwidth = 800,width=0.9\textwidth]{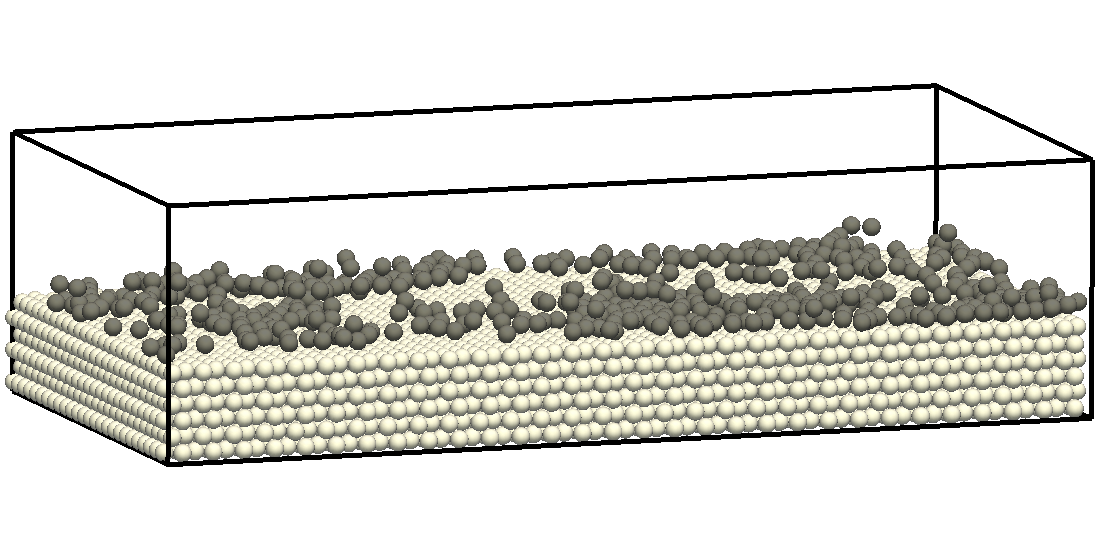}
  \caption{Layout of sediment transport simulation according to~\cite{Kempe14ot}. The white
    particles are fixed on the bottom; the gray particles are movable.}
  \label{fig:layout-st}
\end{figure}

The numerical setup is based on the numerical benchmark studying the motion of 500 movable
particles~\citep{Kempe14ot}. The geometry of the simulation is shown in Figure~\ref{fig:layout-st}.
The dimensions of the domain, the mesh resolutions, and the fluid and particle properties used are
detailed in Table~\ref{tab:param-all}. The CFD mesh in the vertical ($y$-) direction is
progressively refined towards the bottom boundary. Periodic boundary condition is applied in both
$x$- and $z$-directions, no-slip wall condition is applied at the bottom in $y$-direction, and slip
wall condition is applied on the top in $y$-direction. The bulk Reynolds number $Re_b$ of the flow
in the channel is 3010. Six layers of fixed particles are arranged hexagonally to provide a bottom
boundary condition to the moving particles, as is shown in Fig.~\ref{fig:layout-st}. In the
coordinate system, the top of the fixed particle bed is at $y = 0$. To obtain the profiles of fluid
velocity and Reynolds stresses, the simulations are averaged for 50 flow-through times, and spatial
average is performed in the horizontal domain. The bandwidth $b$ used in the averaging procedure is
$2d_p$ in width and thickness directions and $d_p$ in vertical direction.

\begin{figure}[htbp]
  \centering
 \includegraphics[natheight = 500, natwidth = 500,width=0.45\textwidth]{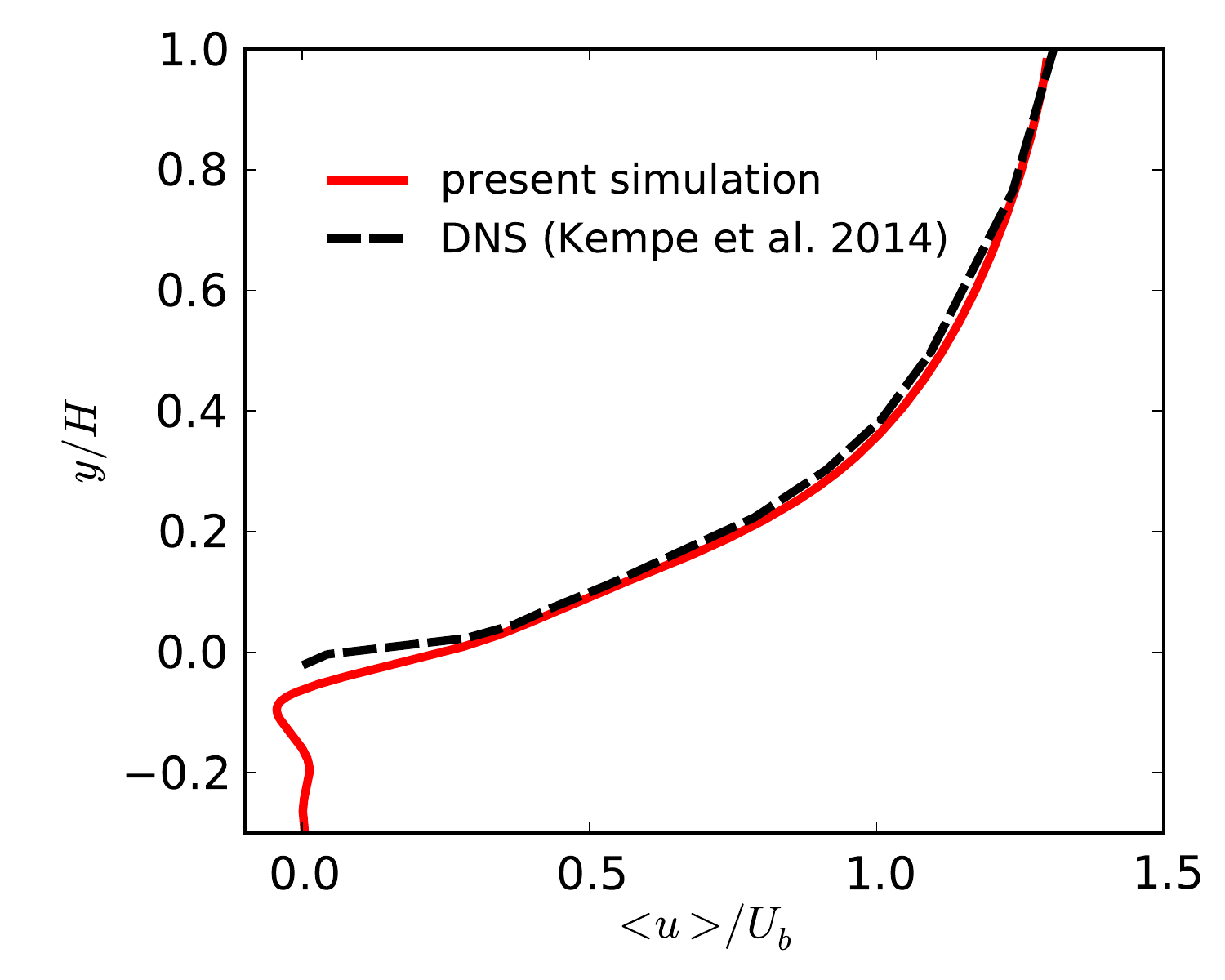}
 \includegraphics[natheight = 500, natwidth = 500,width=0.45\textwidth]{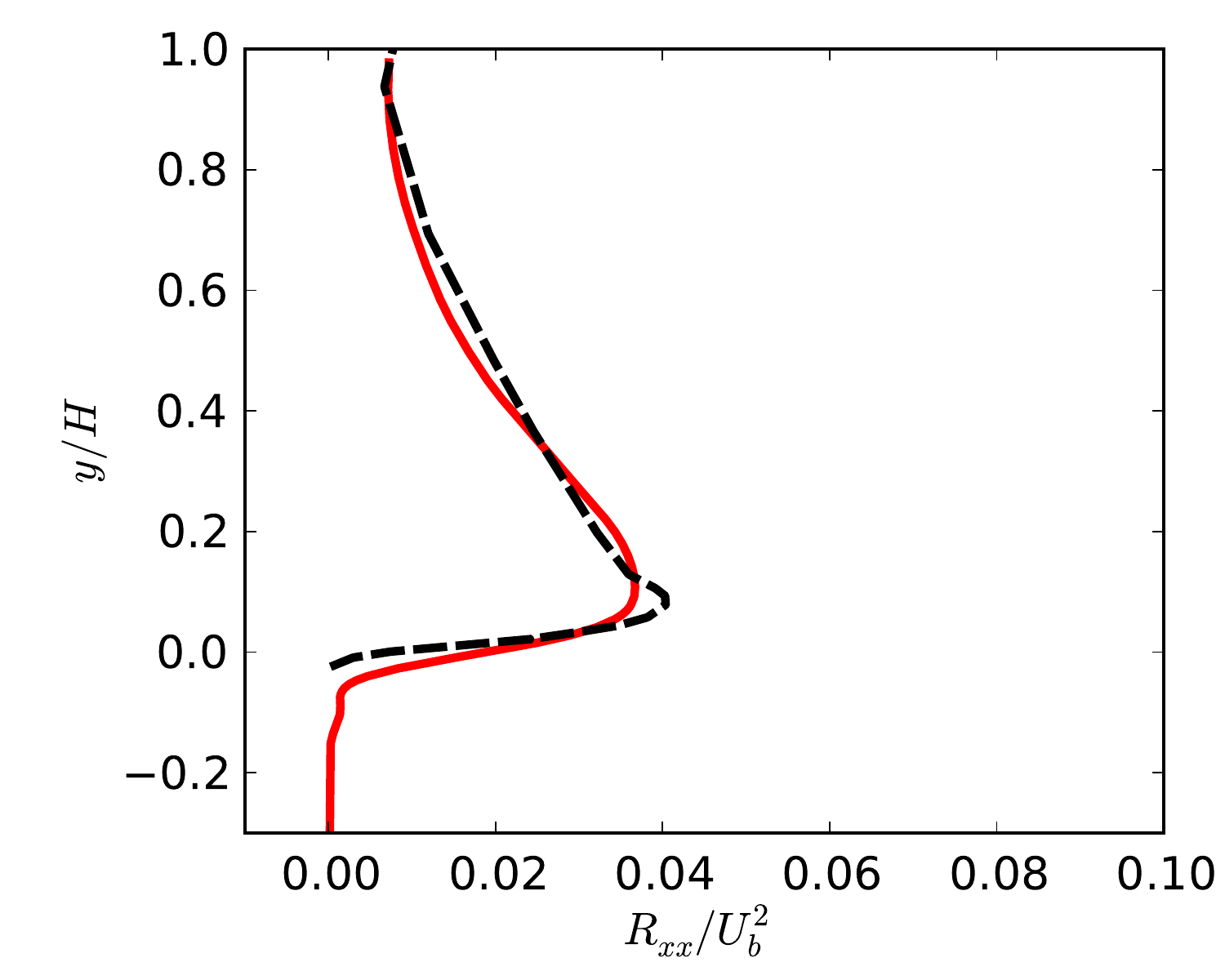}
 \includegraphics[natheight = 500, natwidth = 500,width=0.45\textwidth]{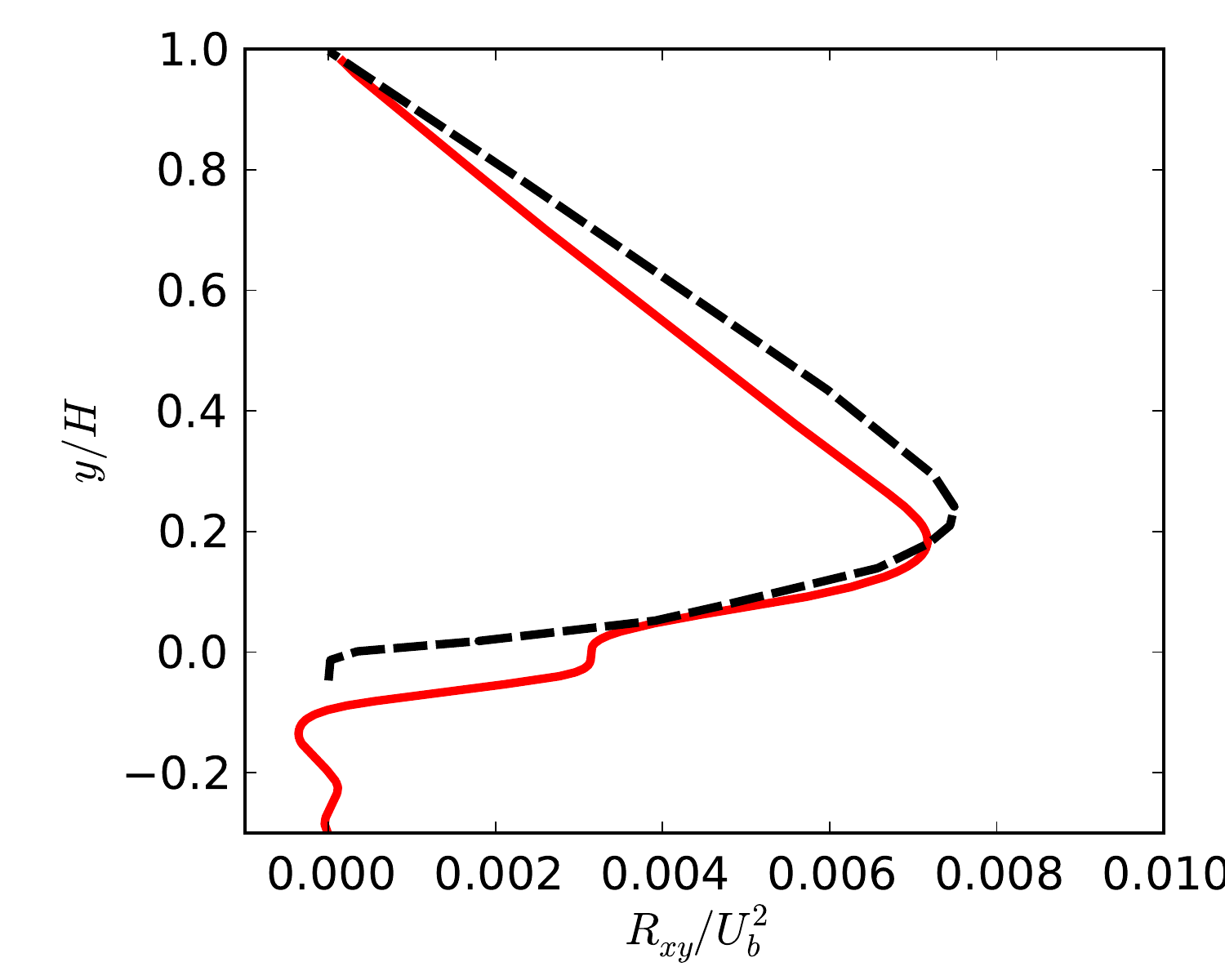}
 \includegraphics[natheight = 500, natwidth = 500,width=0.45\textwidth]{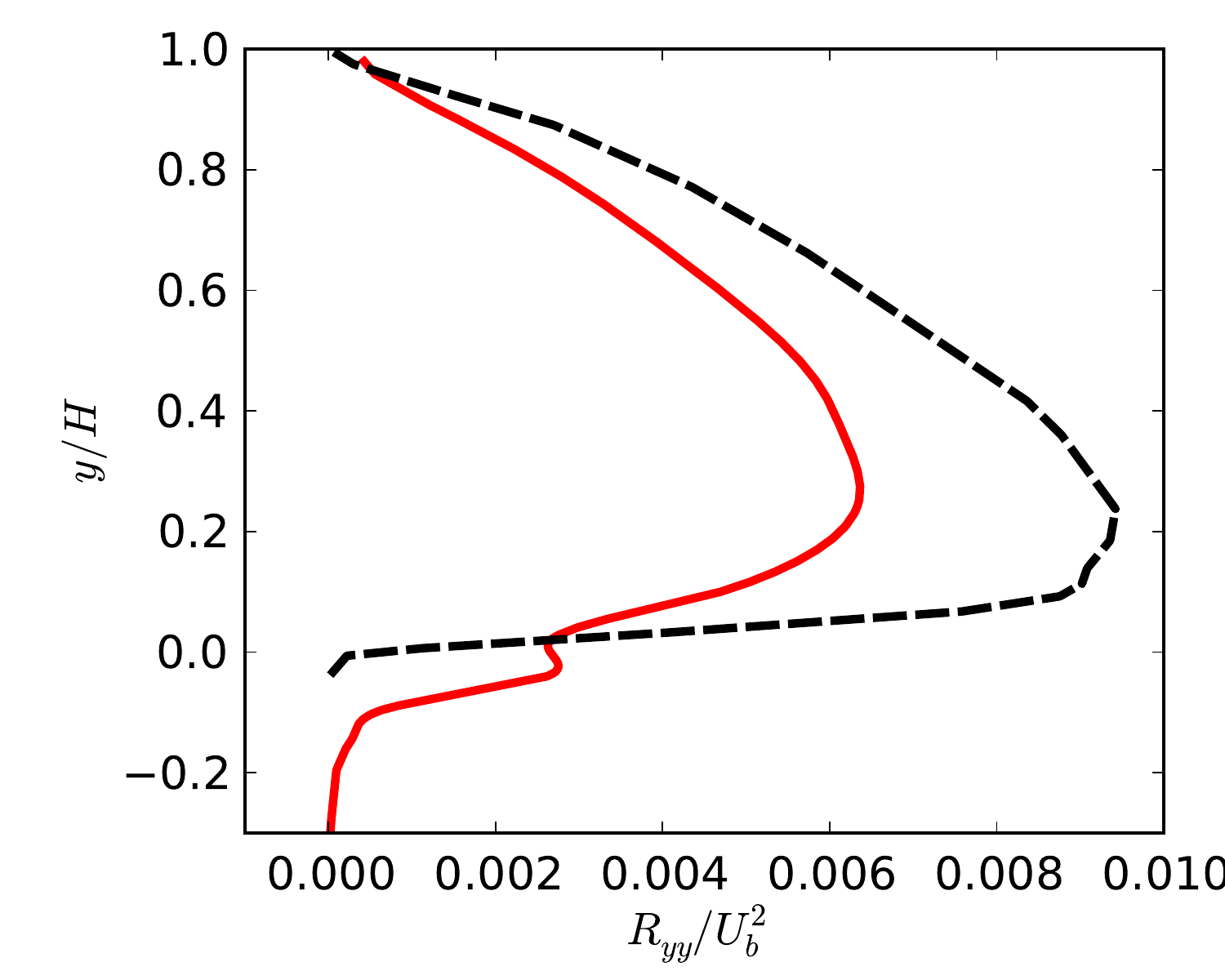}
 \caption{Comparison of mean velocity $\langle u \rangle$ and Reynolds stresses $R_{xx}$, $R_{xy}$,
 and $R_{yy}$ along the wall normal direction ($y$) obtained by using \textit{SediFoam} with the DNS
 results~\citep{Kempe14ot}. The mean location of the particle bed is at $y = 0$.}
  \label{fig:kempe-flow-all}
\end{figure}

The flow properties are presented in Fig.~\ref{fig:kempe-flow-all}, including the streamwise
velocity and Reynolds stresses $R_{xx}$, $R_{xy}$, and $R_{yy}$. Fig.~\ref{fig:kempe-flow-all}(a)
shows that \textit{SediFoam} is able to capture the decrease of the fluid velocity in the near-wall
boundary layer and within the bed, which is due to the drag force of the sediment particles. The
velocity profile near the particle bed at $y = 0$ is negative because of the diffusion effect of the
averaging algorithm.  Therefore, the diffusion bandwidth in the vertical direction is taken as small
as $d_p$ to reduce the effect of the numerical diffusion. It can be seen from
Fig.~\ref{fig:kempe-flow-all}(b) to Fig.~\ref{fig:kempe-flow-all}(d) that the predictions of
different components of Reynolds stresses from the present solver agree well with the numerical
benchmark.  Compared with the flow that has no particles in the channel, \textit{SediFoam} captures
the increase of the Reynolds stresses induced by the motion and collision of particles.
{\color{black}It can be seen that the overall agreement of results obtained by using \textit{SediFoam}
and DNS is good for $R_{xx}$ and $R_{yy}$. The discrepancy in $R_{yy}$ is slightly larger than other
components, which may be attributed to the fact that CFD--DEM do not resolve the flow fluctuation at
the particle surface and cannot capture this quantity as good as the DNS.} 

\begin{figure}[htbp]
 \centering
 \includegraphics[natheight = 500, natwidth = 500,width=0.45\textwidth]{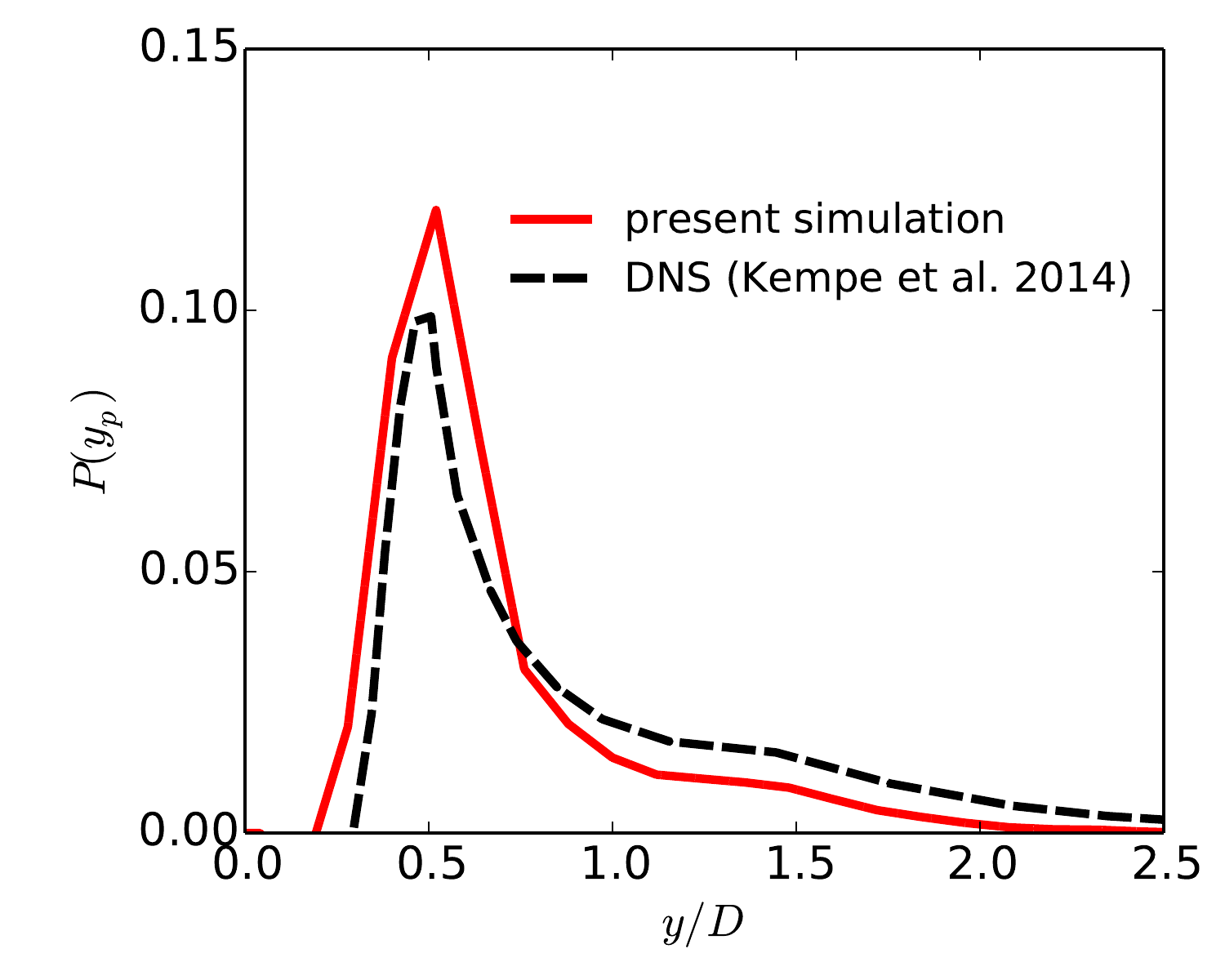}
 \includegraphics[natheight = 500, natwidth = 500,width=0.45\textwidth]{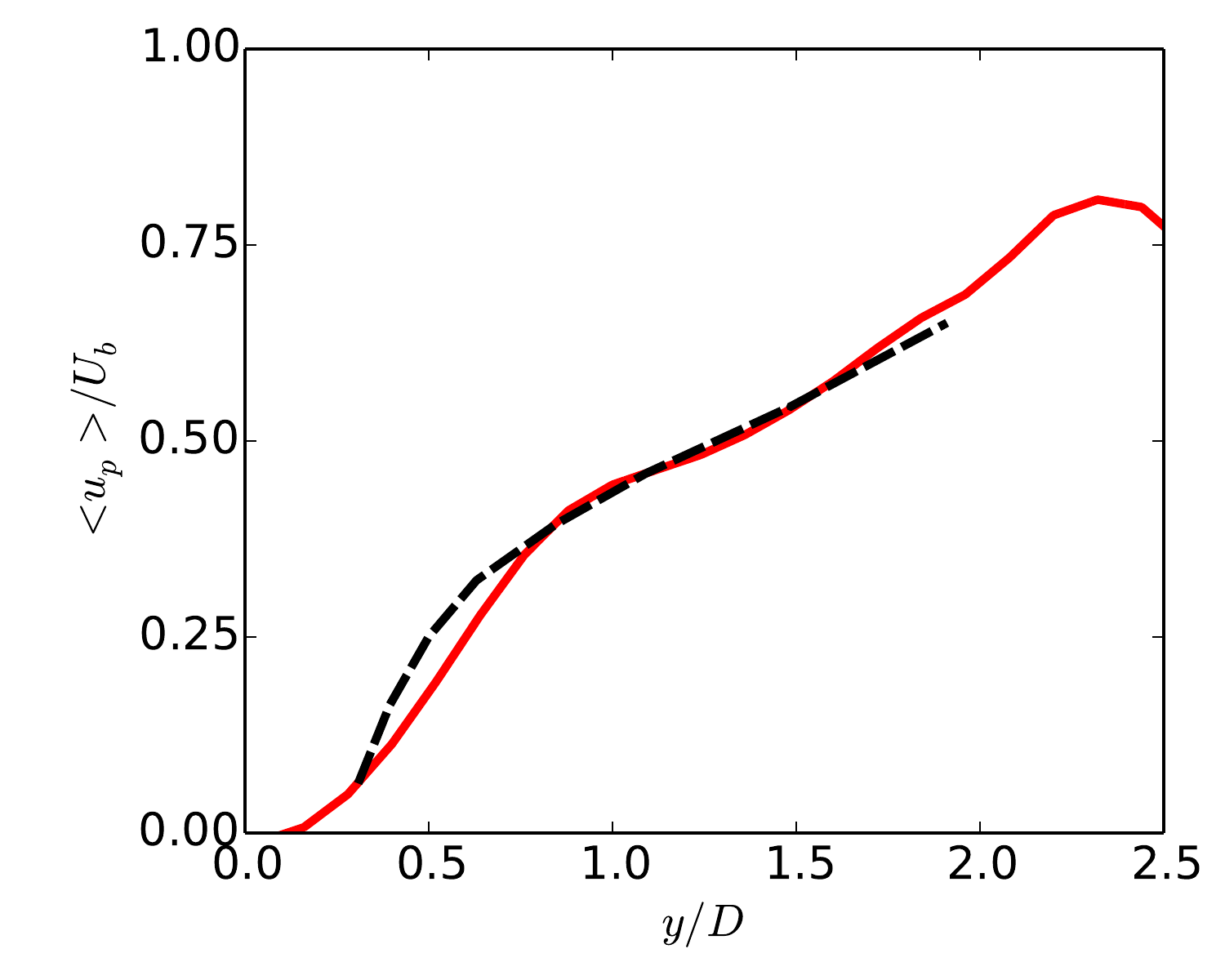}
 \caption{Probability density function and streamwise velocity in the simulation of 500 movable
   particles.}
 \label{fig:kempe-sedi-all}
\end{figure}

Other quantities of interest in sediment transport include solid volume fraction and particle
velocity. Figure~\ref{fig:kempe-sedi-all} demonstrates the time-averaged probability density
function (PDF) and streamwise velocity of the particles at different vertical locations. It can be
seen that both the probability density function and streamwise velocity agree well with the results
in the numerical benchmark~\citep{Kempe14ot}. In the present simulation, the averaged streamwise
velocity of all particles $u_p/u_b$ is 0.28. Compared with the prediction 0.35 from the numerical
benchmark, the error is 20\%. However, this error is insignificant since $u_p$ is proportional to
the sediment transport rate, which varies significantly in the experimental measurements. Therefore,
the accuracy of CFD--DEM modeling is acceptable. It is noteworthy considering the fact that the
total computational costs of CFD--DEM are much smaller than the interface-resolved method. For this
case, the number of CFD mesh in the present simulation is $1.4\times10^5$, whereas this number in
the DNS simulation is $1.0\times10^8$. 

\subsection{Case 3: Sediment Transport with $O(10^5)$ Particles}
\label{sec:large-sediment}
The purpose of this simulation is to test the performance of the present code in the simulation of
larger problems of sediment transport. The results of both bed load sediment transport and suspended
sediment transport are demonstrated. 

The layout of this case is similar to Case 2. The domain geometry, the mesh resolution, and the
properties of fluid and particles are detailed in Table~\ref{tab:param-all}. Periodic boundary
condition is applied in both $x$- and $z$-directions. Slip wall condition is applied at the top of
the domain, and no-slip wall condition is applied at the bottom. Three layers of solid particles are
fixed at the bottom to provide rough wall boundary condition for DEM simulation. The flow
velocities in five numerical simulations range from $0.3$~m/s to $1.1$~m/s. Each simulation is
performed for 50 flow-through times for time-averaging. The bandwidth $b$ used in the averaging
procedure is $4d_p$ in width and thickness directions and $2d_p$ in vertical direction. The geometry
of this case is identical to the numerical tests using the same number of particles
by~\citet{schmeeckle14ns}. 

\begin{figure}[htbp]
  \centering
  \subfloat[Distribution and velocity of sediment particles]{
      \label{fig:sedi-velo}
      \includegraphics[natheight = 500, natwidth = 1000,width=0.6\textwidth]{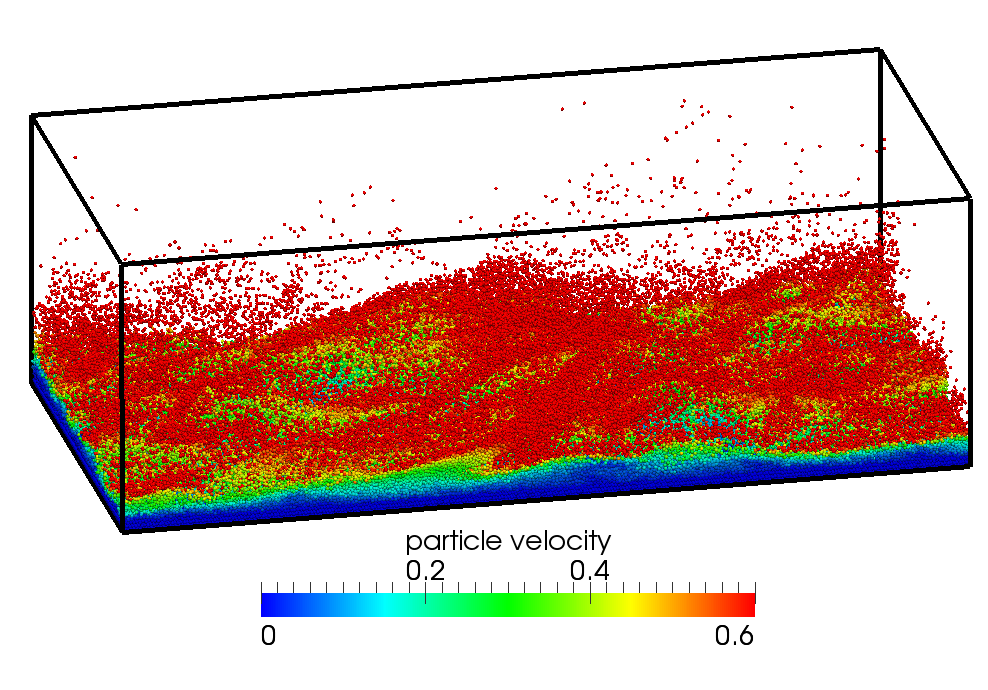}
    }
    \vspace{0.001\textwidth}
    \subfloat[Sediment transport rate]{
      \label{fig:sedi-rate}
      \includegraphics[natheight = 500, natwidth = 500,width=0.45\textwidth]{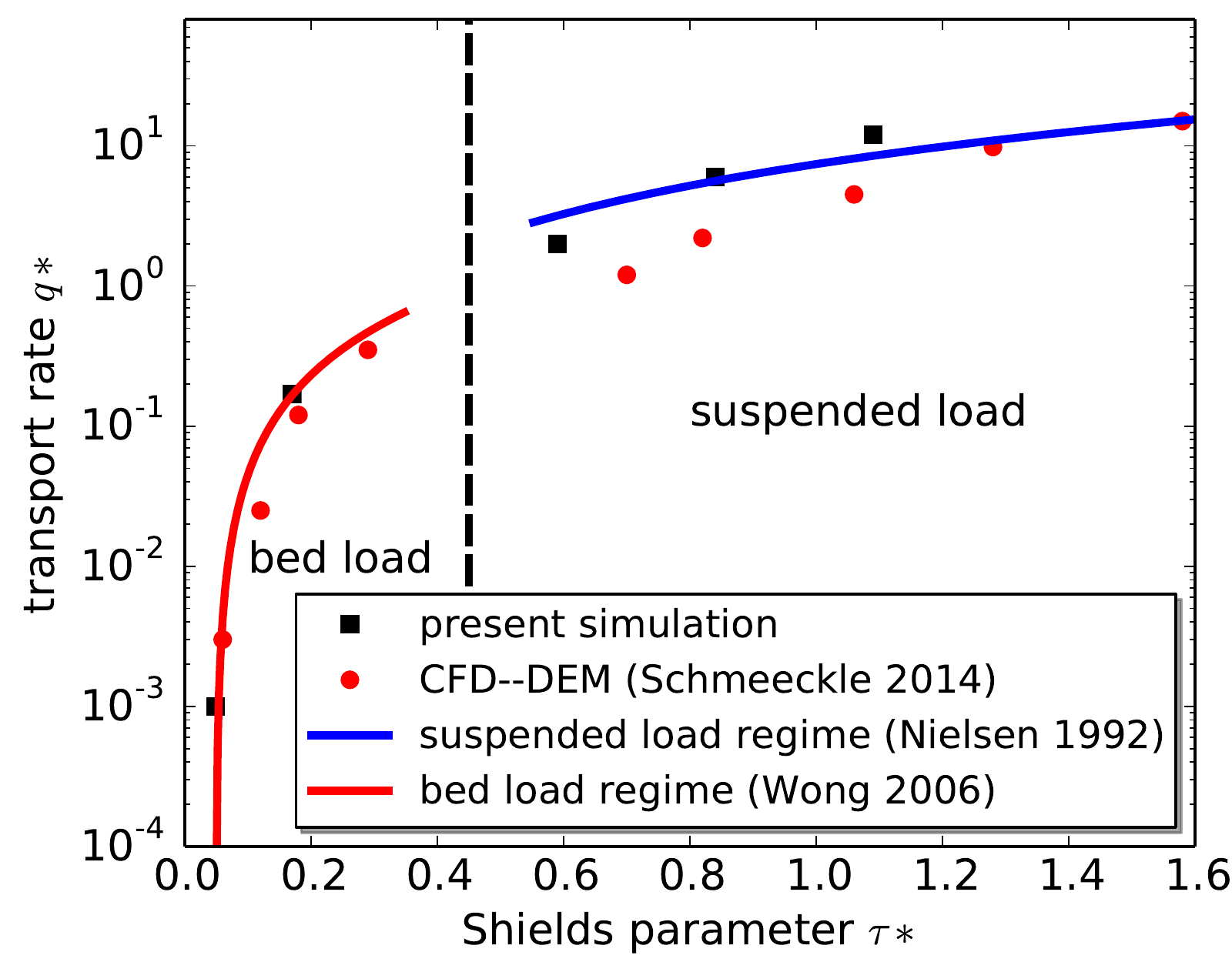}
    }
    \hspace{0.001\textwidth}
    \subfloat[Surface friction]{
      \label{fig:sedi-cf}
      \includegraphics[natheight = 500, natwidth = 500,width=0.45\textwidth]{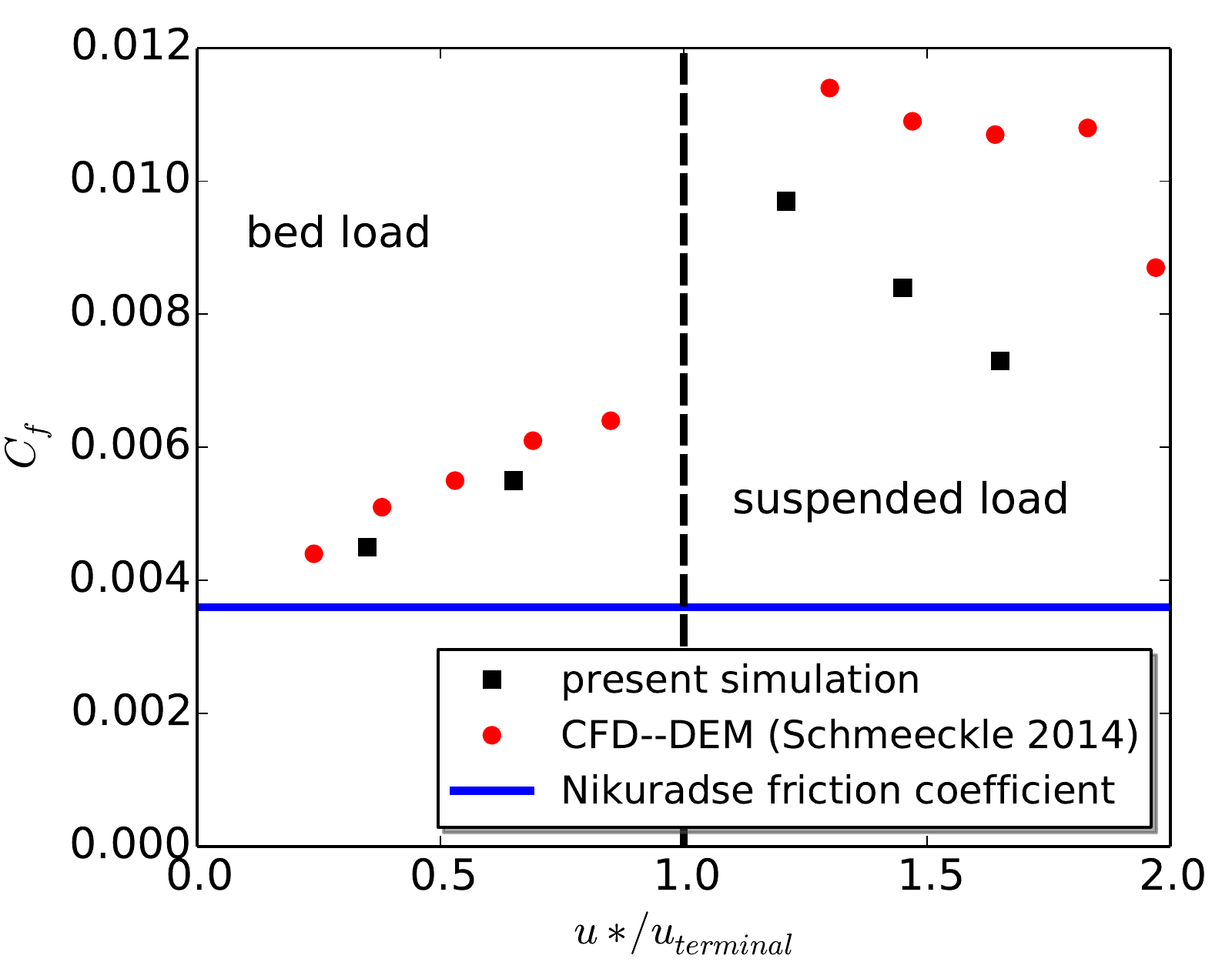}
    }
    \caption{CFD--DEM simulation of sediment transport in a periodic channel with \textit{SediFoam},
    using 330,000 particles. The snapshot of the particles (colored by their velocity magnitudes)
    and comparison with empirical formulas of bed load transport and suspended load in different
    regimes~\citep{wong06ra,nielsen92cb} are displayed, showing favorable agreement.}
  \label{fig:sedi2-rate}
\end{figure}

The averaged properties of sediment particles are presented in Fig.~\ref{fig:sedi2-rate}, including
the sediment transport rate and the friction coefficient. The sediment transport rate $q_{sx}$ is
obtained by multiplying the mean streamwise velocity of all particles by the total volume of the
particles, and divide it by the area of the horizontal plane. The non-dimensional sediment fluxes
$q_* = q_{sx}/((\rho_s/\rho_f - 1)g d_p^3)^{1/2}$ at different Shields parameters $\tau_* = \rho_f
u_*^2/(\rho_s - \rho_f)g d_p$ are shown in Fig.~\ref{fig:sedi2-rate}(b). It can be seen in the
figure that the sediment transport rates agree favorably with the experimental data. Note that the
$q_*$ in different regimes (i.e., bed load and suspended load) used different regression curves. It
is worth mentioning that the predictions of $q_*$ in suspended load regime using \textit{SediFoam}
are better than the results obtained in the literature~\citep{schmeeckle14ns}. This is because the
drag force model in the present simulations can better predict the motion of sediment particles
since the influence of $\varepsilon_s$ is accounted for. The coefficient of friction of the surface
is defined as $C_f = u_*^2/\langle u \rangle^2$ and describes the roughness of the bed. Shown in
Fig.~\ref{fig:sedi2-rate}(c), $C_f$ obtained in the present simulation are larger than the
predictions by the law of the wall (i.e., Nikuradse value).  This increase in $C_f$ is because the
hydraulic roughness over a loose bed is larger in the presence of movable particles. Note that the
$C_f$ predicted by \textit{SediFoam} is slightly smaller than the results predicted
by~\cite{schmeeckle14ns}. This is because \cite{schmeeckle14ns} ignored the volume fraction in the
numerical simulation and used a smaller $\langle u \rangle$ when calculating $C_f$.

\begin{figure}[htbp]
  \centering
    \subfloat[Solid volume fraction $\varepsilon_s$]{
      \label{fig:sedi-uMean}
      \includegraphics[natheight = 500, natwidth = 400,width=0.45\textwidth]{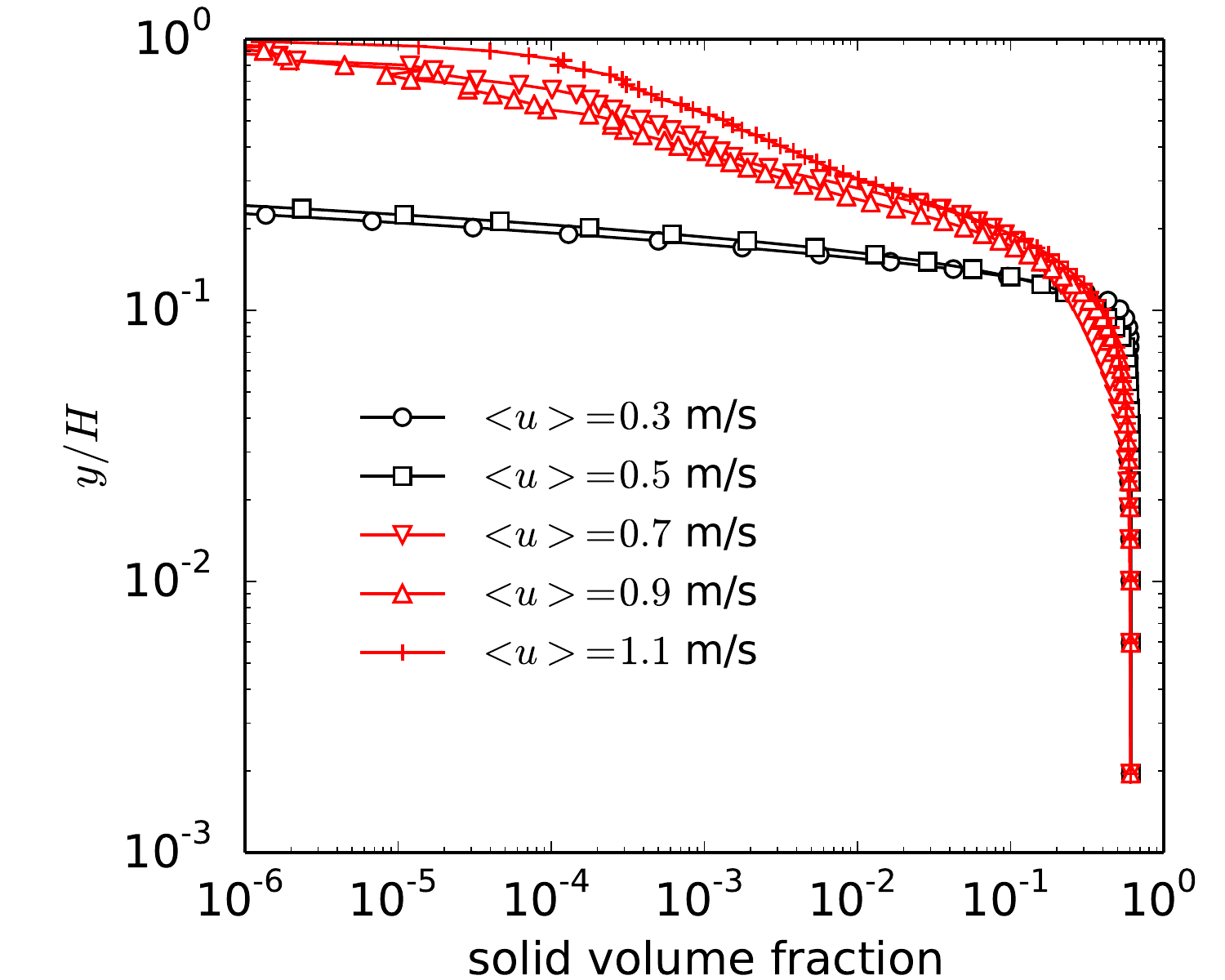}
    }
    \vspace{0.001\textwidth}
    \subfloat[Streamwise flow velocity]{
      \label{fig:sedi-c}
      \includegraphics[natheight = 500, natwidth = 400,width=0.45\textwidth]{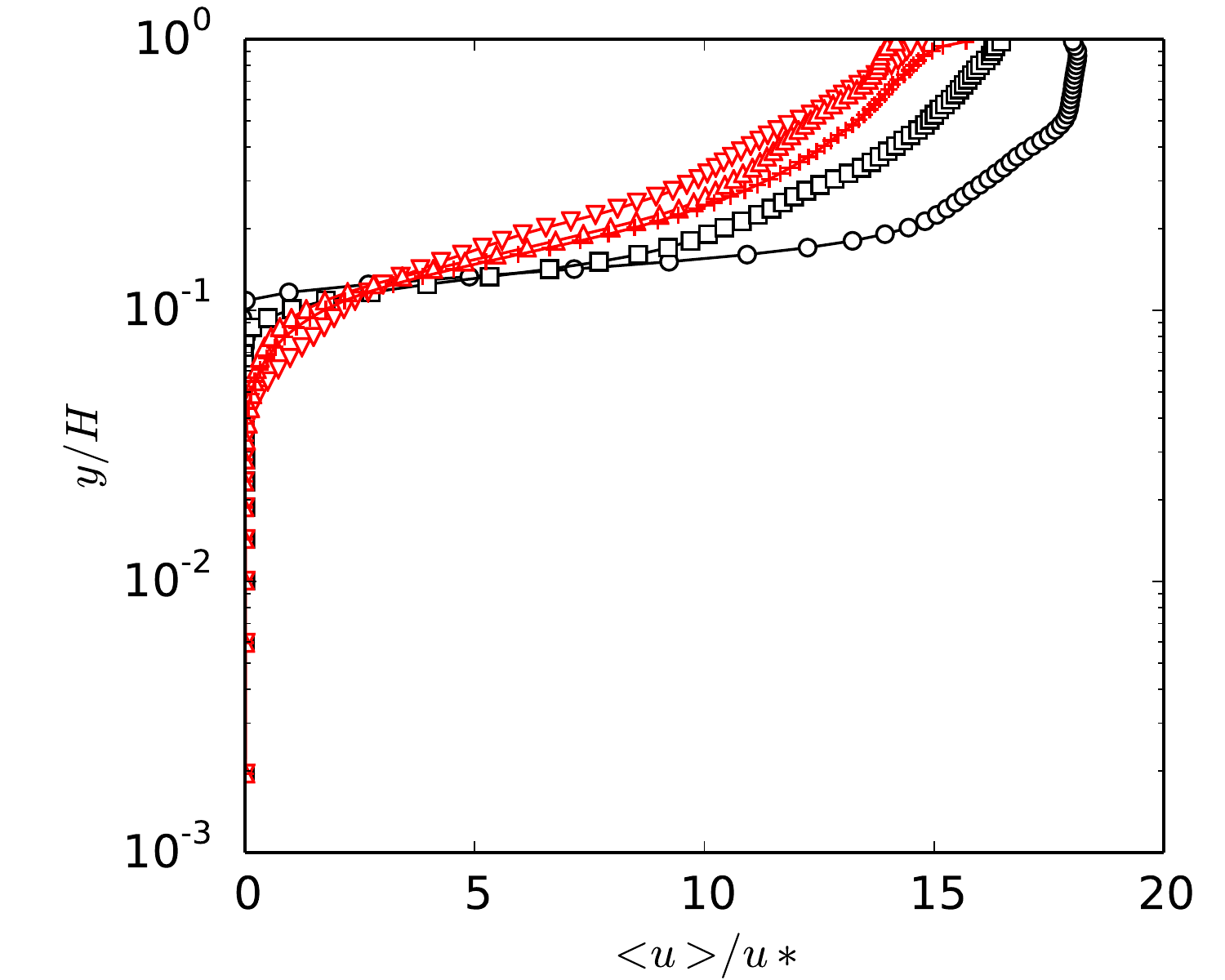}
    }
    \caption{The solid volume fraction $\varepsilon_s$ and streamwise flow velocity of
    different flow velocities. The blue lines are the simulations in bed load regime; the red lines
    are the simulations in suspended load regime.}

  \label{fig:sedi2-fluid}
\end{figure}

The spatially and temporally averaged profiles of solid volume fraction and flow velocity are shown
in Fig.~\ref{fig:sedi2-fluid}. It can be seen from Fig.~\ref{fig:sedi2-fluid}(a) that the
$\varepsilon_s$ above $y/H = 0.2$ is approximately zero in the bed load regime. This is because the
sediment particles are rolling and sliding on the sediment bed in this regime. In the suspended load
regime, the particles are suspended in the flow due to turbulent eddies.  Therefore, the
$\varepsilon_s$ above $y/H = 0.2$ is much larger. The flow velocity profiles also vary at different
regimes of sediment transport. In the bed load regime, the height of the sediment bed is
approximately $0.1H$. The sediment particles on the sediment bed are moving slowly in this regime.
Hence, the streamwise flow velocity decays rapidly under the sediment bed. In the suspended load
regime, the flow velocity at the sediment bed is more diffusive since the particles at the sediment
bed are moving more rapidly. Compared with the data obtained in the
literature~\citep{schmeeckle14ns}, the overall agreement of solid volume fraction and flow velocity
is satisfactory.

\section{Scalability}
\label{sec:parallel-eff}
The parallel efficiency is crucial to CFD--DEM solvers for simulations of large and complex
problems. In Section~\ref{sec:parallel-fb} and~\ref{sec:parallel-st}, the parallel efficiencies in
the simulations of fluidized bed and sediment transport are studied separately. This is because the
setup, the flow regime, and the behavior of particles are different between the two cases. The CPU
time spent on different parts of \textit{SediFoam} is detailed in Section~\ref{sec:parallel-comp}. 

Both strong and weak scalability are studied in the parallel efficiency tests. The strong
scalability is to evaluate the performance of a solver for a constant sized problem being separated
by using more and more processors. The amount of the total work is constant but the amount of
communication work increases. The parallel efficiency of the strong scalability test is
$N_{p0}t_{p0}/N_{pn}t_{pn}$, where $N_{p0}$ and $N_{pn}$ are the number of processors employed in
the simulation of the baseline case and the test case, respectively; $t_{p0}$ and $t_{pn}$ are the
CPU time. The speed-up is defined as $t_{p0}/t_{pn}$, which is the relative improvement of the CPU
time when solving the problem. The weak scalability evaluates the performance of a solver for the
problem with increasing number of processors. In the weak scalability test, the amount of the work
in different processors is the same. The parallel efficiency of the weak scalability test is
$t_{p0}/t_{pn}$. The scale-up is defined as $N_{pn}t_{p0}/t_{pn}$, which is the improvement in the
scale of the problem that \textit{SediFoam} can solve.

\subsection{Scalability of Fluidized Bed Simulation}
\label{sec:parallel-fb}
The validation case of fluidized bed~\citep{sun15db2} is applied to test the parallel efficiency of
\textit{SediFoam}. PCM-based averaging, which requires little computational cost, is applied to
reduce the influence of averaging algorithm. In the simulation of fluidized bed, various numbers of
processors are used from 4 to 256. The computational hours are calculated by running 100 time steps
in each case. 

\begin{figure}[!htbp]
  \centering
    \subfloat[strong scaling]{
      \label{fig:fb-s}
      \includegraphics[natheight = 500, natwidth = 400,width=0.45\textwidth]{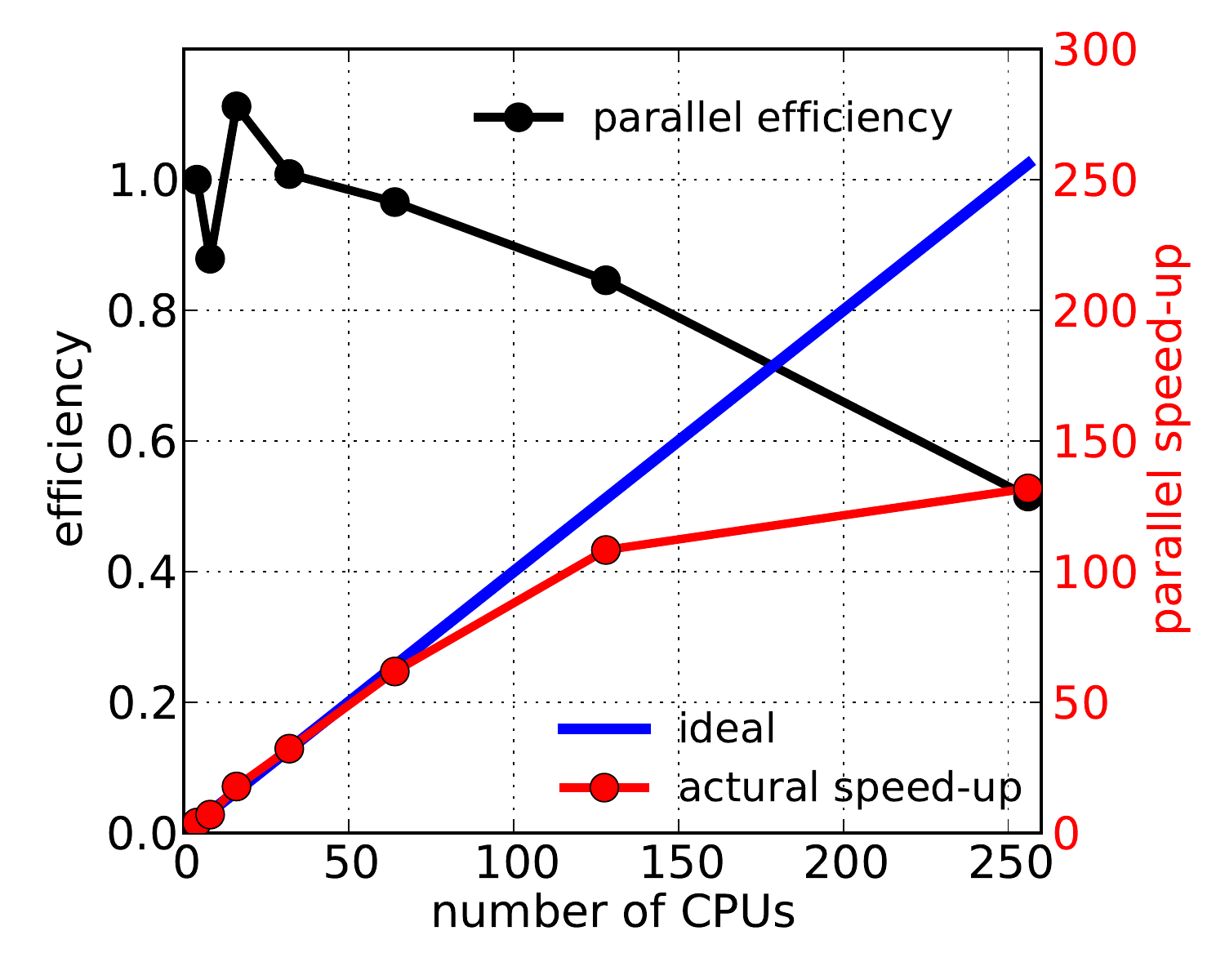}
    }
    \subfloat[weak scaling]{
      \label{fig:fb-w}
      \includegraphics[natheight = 500, natwidth = 400,width=0.45\textwidth]{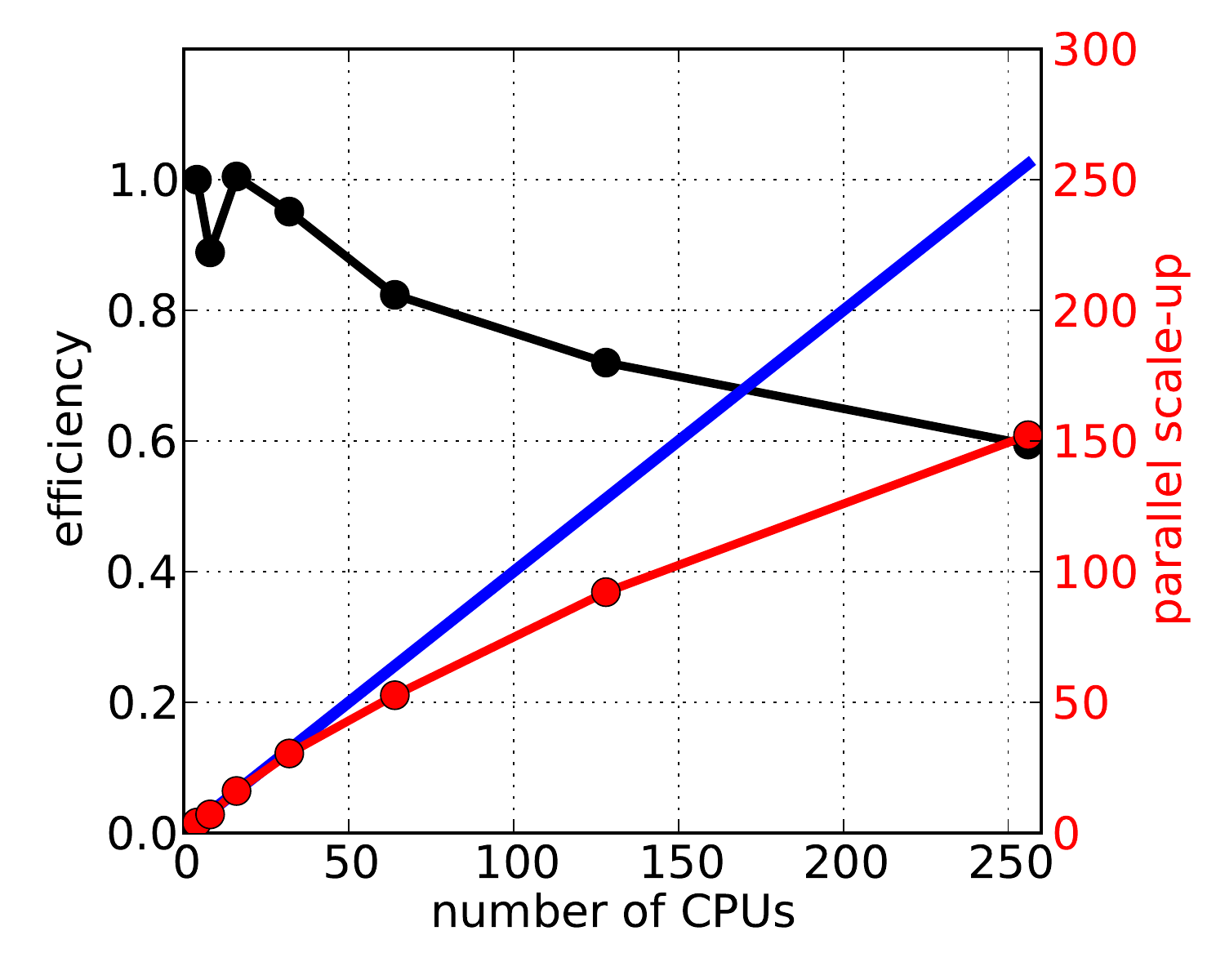}
    }
  \caption{The parallel efficiency of \textit{SediFoam} for strong scaling and weak scaling in
  fluidized bed. The test case employs up to 21 million sediment particles
  using up to 256 CPU cores. The simulation with the smallest number cores is regarded as base case
  when computing speed-up. Tests are performed on Virginia Tech's BlueRidge cluster.
  (http://www.arc.vt.edu/)}
  \label{fig:paraEfficiency-fb}
\end{figure}

To test the strong scalability of the code for large-scale simulations, the numerical test is
performed using 5.3 million sediment particles and 1.3 million CFD cells. The domain is $1056$~mm
$\times~600$~mm $\times~240$~mm in width ($x$-), height ($y$-), and transverse thickness ($z$-)
directions. The results obtained in the strong scalability test are shown in
Fig.~\ref{fig:paraEfficiency-fb}(a). It can be seen that the parallel efficiency of
\textit{SediFoam} is close to 100\% when employing less than 32 processors and is still as high as
85\% when using 128 processors. However, the parallel efficiency decreases to 52\% when using 256
processors. In the test using 256 processors, the number of sediment particles in each processor is
as small as 20,000. From the results reported in the literature, the parallel efficiency of
\textit{SediFoam} is approximately 15\% higher than other solvers when using this number of
particles~\citep{gopalakrishnan13dop}. {\color{black} It is noted in
Fig.~\ref{fig:paraEfficiency-fb}(a) that the parallel efficiency fluctuates when using 4 to 32
processors. This may attributes to the fluctuation of the CPU time when solving the linear system.
When using different numbers of processors, the decomposition of the domain leads to different
linear systems and different converge rates when solving these linear systems. This fluctuation may
also be due to the variation of CPU cache or memory. The fluctuation of the CPU time can also be
seen in the speed-up curve, but the magnitude is small.}

To test the weak scalability of \textit{SediFoam}, $8.3\times10^4$ particles are located in each
processor with the dimensions of $66$~mm $\times~600$~mm $\times~60$~mm in width ($x$-), height
($y$-), and transverse thickness ($z$-) directions. The numbers of processors vary from 4 to 256,
and thus the total numbers of particles vary from 0.3 million to 21 million. It can be seen in
Fig.~\ref{fig:paraEfficiency-fb}(b) that good weak scalability is observed. The parallel efficiency
is close to 100\% when using less than 32 processors and gradually decreases to 61\% when using 256
processors. Therefore, the ultimate scale-up of the code by using 256 processors is as large as 156.

\subsection{Scalability of Sediment Transport Simulation}
\label{sec:parallel-st}
Simulations of sediment transport cases are performed to test the parallel efficiency of
\textit{SediFoam} when the number of CFD mesh is larger than the number of particles. The
diffusion-based averaging algorithm is applied in the simulations. The numbers of processors used
vary from 8 to 512.

\begin{figure}[!htbp]
  \centering
    \subfloat[strong scaling]{
      \label{fig:st-s}
      \includegraphics[natheight = 500, natwidth = 400,width=0.45\textwidth]{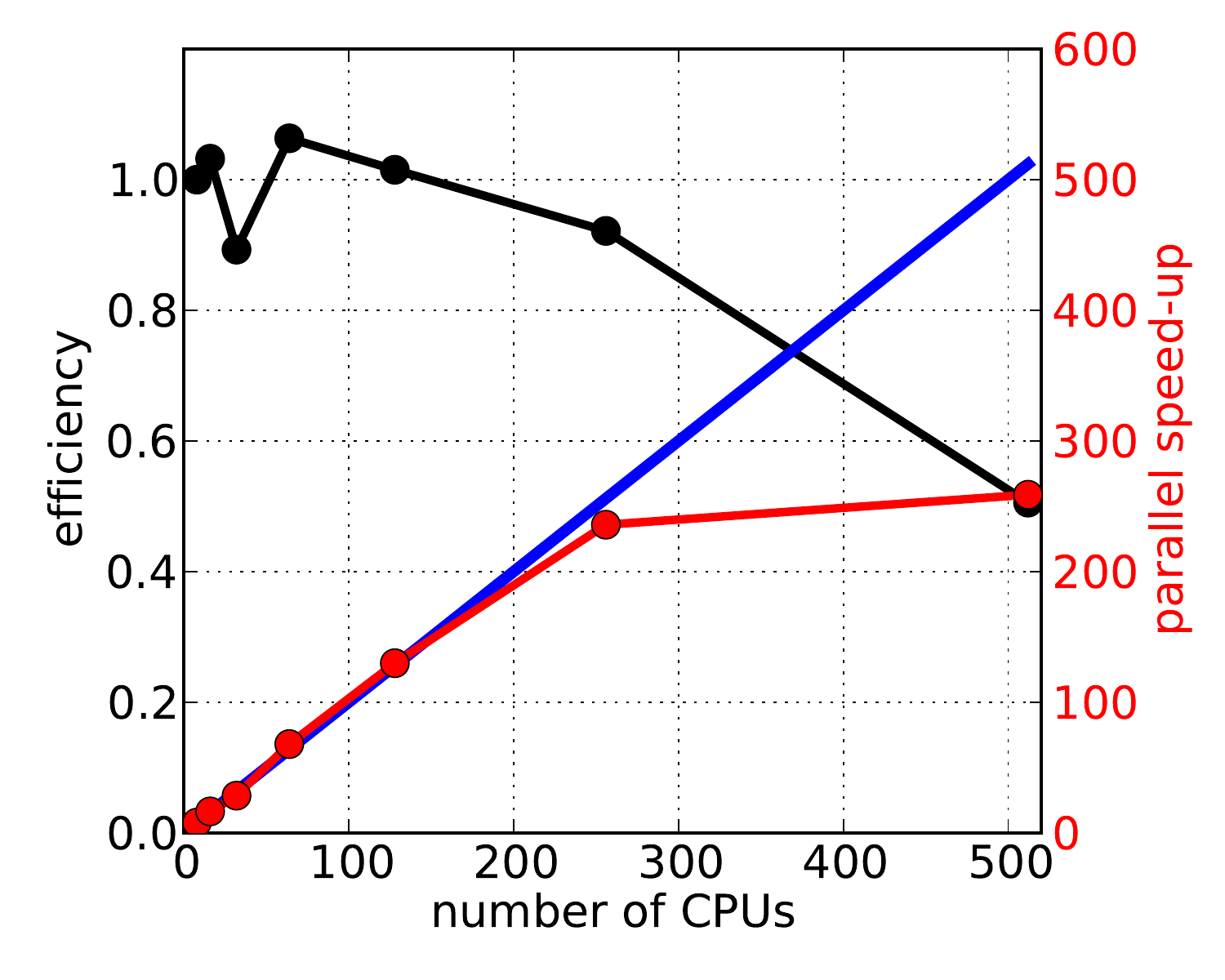}
    }
    \subfloat[weak scaling]{
      \label{fig:st-w}
      \includegraphics[natheight = 500, natwidth = 400,width=0.45\textwidth]{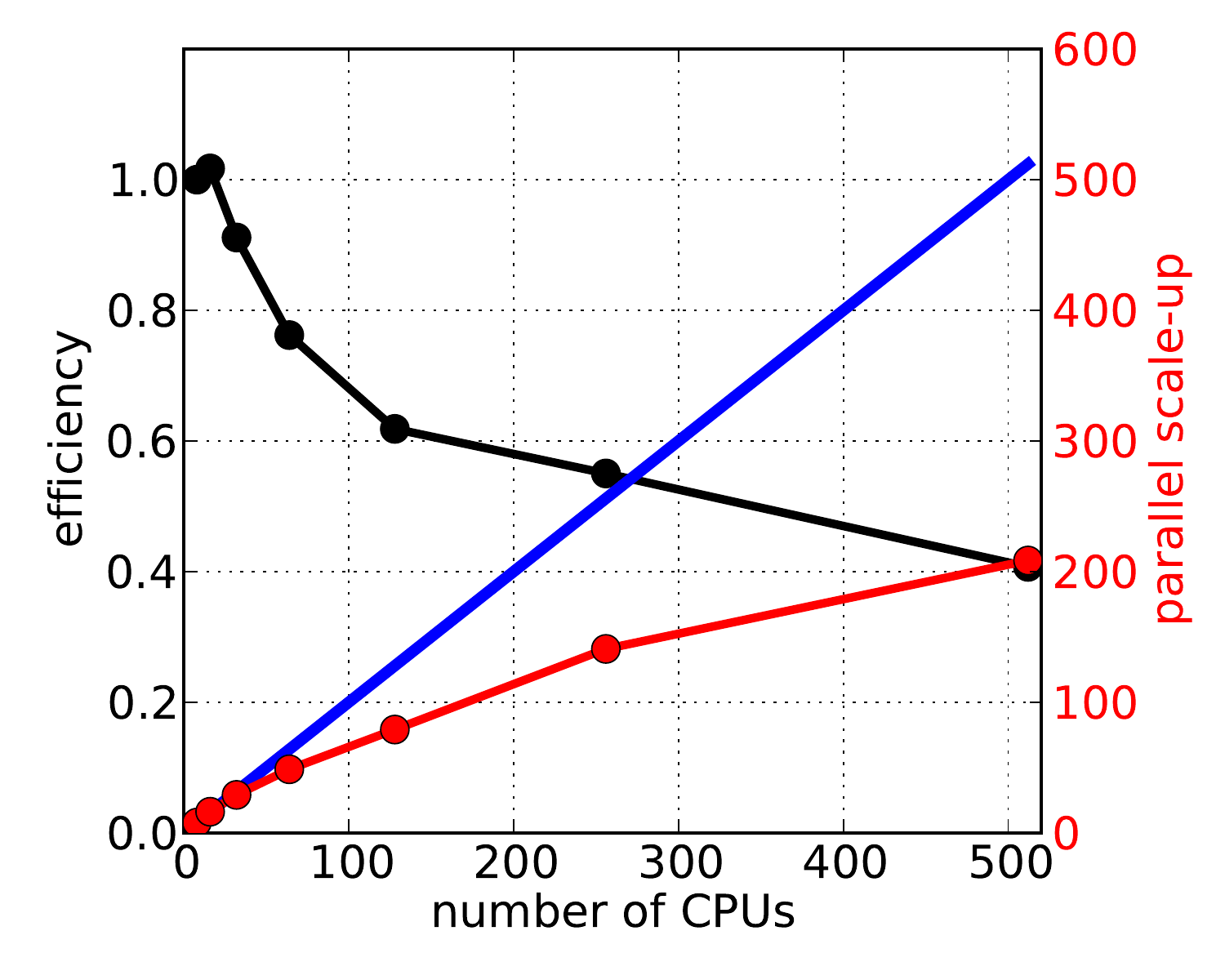}
    }
  \caption{The parallel efficiency of \textit{SediFoam} for strong scaling and weak scaling in
sediment transport. The test case employs up to 40 million sediment particles using up to 512 CPU
cores. The simulation with the smallest number cores is regarded as base case when computing
speed-up.}
  \label{fig:paraEfficiency-st}
\end{figure}

To test the strong scalability of the code for large-scale simulations, we expanded the domain in
Case 3 to be $480$~mm $\times~40$~mm $\times~480$~mm in width ($x$-), height ($y$-), and transverse
thickness ($z$-) directions. The numbers of sediment particles and CFD cells are 10 million and 13
million, respectively. It can be seen in Fig.~\ref{fig:paraEfficiency-st}(a) that good scalability
and parallel efficiency are obtained in this test. The parallel efficiency is close to 100\% when
using less than 64 processors and decreases to 52\% when using 512 processors. 

To test the weak scalability for sediment transport, $7.8\times10^4$ particles are located in each
processor with the dimension of $60$~mm $\times~40$~mm $\times~30$~mm in width ($x$-), height
($y$-), and transverse thickness ($z$-) directions. The total numbers of particles vary from 0.6
million to 40 million. The parallel efficiency and the scale-up of \textit{SediFoam} are shown in
Fig.~\ref{fig:paraEfficiency-fb}(b). Compared with the results obtained in the fluidized bed test,
the parallel efficiency in sediment transport test is lower. This is because solving the fluid
equations is difficult in parallel when the number of CFD cells is very large.

In summary, although the parallel efficiency of \textit{SediFoam} in the sediment transport cases is
smaller than the fluidized bed cases, the general efficiency is still satisfactory compared to other
CFD--DEM solvers~\citep{gopalakrishnan13dop,amritkar12op,capecelatro13ae}. Moreover, simulations of
$O(10^7)$ number of particles are performed, which demonstrates the capability of \textit{SediFoam}
in the simulation of relatively large scale problems.

\subsection{Parallel performance of different components}
\label{sec:parallel-comp}
The investigation of the computational costs of different components in \textit{SediFoam} is
detailed in this section. The study consists of both simulations of fluidized bed and sediment
transport, using the results obtained in the tests of Section~\ref{sec:parallel-fb}
and~\ref{sec:parallel-st}. The time spent on the CFD module, the DEM module, the coupling
between the two modules, and the averaging algorithm are discussed.

The results obtained in the fluidized bed simulation are shown in Table.~\ref{tab:weight-fb}. It can
be seen in the table that the most time-consuming process is solving the collision between the
particles, which accounts for 46\% to 76\% of the total costs. Solving the fluid equations is the
second most time-consuming process, which takes 21\% to 31\% of the total computational costs for
both strong scaling and weak scaling. It is noted that the coupling process takes less than 2\%
of the total costs when using less than 64 processors.  However, the proportion of this part
increases significantly to about 20\% when using 256 processors. This is because the time spent on
the inter-processor communication increases due to the increase number of processors used in the
simulation. This increase can also be seen in the weak scaling test for both the CFD and DEM
modules, which are well-established solvers with good parallel efficiency.

The computational costs of different components in the sediment transport simulation are detailed in
Table.~\ref{tab:weight-st}. Since the number of CFD cells is larger than the number of sediment
particles, the computational costs of the CFD module are larger than the DEM module. It can be seen
that the proportion of CFD part is approximately 21\% to 48\% the total computational costs for both
strong scaling and weak scaling tests. The DEM part accounts for about 17\% to 37\% of the total
costs. It is noted that the computational overhead of the averaging also accounts for 20\% to 33\%
of the total computational overhead. The time spent on the coupling is less than 2\% of the total
time when using less than 128 processors. However, when using as many as 512 processors the time
spend on the communication increases significantly.

\section{Conclusion}
\label{sec:conclude}
In this work, the parallelized open-source CFD--DEM solver \textit{SediFoam} is developed with
emphasis on the simulation of sediment transport. The CFD and DEM modules are based on OpenFOAM and
LAMMPS, respectively. The communication between the modules is implemented using parallel algorithm
to enable the simulation of large-scale problems.

Numerical validations are performed to test the capability of the present solver. The single
particle sedimentation test demonstrates the importance of added mass and lubrication in CFD--DEM
simulations. In the numerical simulations of with 500 sediment particles, the fluid and particle
properties obtained are consistent with the results obtained by interface-resolved method. This
indicates the accuracy of \textit{SediFoam} is desirable. The numerical simulation using $O(10^5)$
particles demonstrates the capability of \textit{SediFoam} in the simulation of various regimes in
sediment transport.

Parallel efficiency tests are conducted to investigate the scalability of \textit{SediFoam}. From
the test, the scalability of \textit{SediFoam} is satisfactory compared with other existing CFD--DEM
solvers. This demonstrates that \textit{SediFoam} is a desirable solver for the simulation of
sediment transport of large-scale and complex problems.

\section{Acknowledgment}

The computational resources used for this project are provided by the Advanced Research Computing
(ARC) of Virginia Tech, which is gratefully acknowledged. We also thank Dr. Jin Sun at University of
Edingburgh, who helped improve the quality of the manuscript through technical discussions.
Moreover, we thank Dr. Calantoni for the discussions, which helped with the numerical simulations in
the present paper. The authors gratefully acknowledge partial funding of graduate research
assistantship from the Institute for Critical Technology and Applied Science (ICTAS, grant number
175258) in this effort.

\bibliographystyle{elsarticle-harv}

\begin{table}[!htbp]
  \caption{Parameters of the numerical simulations.}
 \begin{center}
 \begin{tabular}{lcccc}
   \hline
                                                & Case 1    & Case 2    & Case 3 \\
   \hline
   domain dimensions                            &\\
   \qquad width $(L_x)$~(mm)                    & 100   & 58.4  & 120 \\
   \qquad height $(L_y)$~(mm)                   & 200   & 13.5  & 40 \\
   \qquad transverse thickness $(L_z)$~(mm)     & 100   & 29.2  & 60 \\
   mesh resolutions                             &\\
   \qquad width $(N_x)$                         & 10    & 52    & 140 \\
   \qquad height $(N_y)$                        & 20    & 102   & 65 \\
   \qquad transverse thickness $(N_z)$          & 10    & 26    & 60 \\
   particle properties & \\
   \qquad total number                          & 1     & 500   & $3.3\times10^5$ \\
   \qquad diameter $d_p$~(mm)                   & 6     & 1.12  & 0.5 \\
   \qquad density $\rho_s$~($\times 10^3~\mathrm{kg/m^3}$) 
          & 2.5/7.8 (sand/steel)  & 2.0  & 2.5 \\
   \qquad particle stiffness coefficient~(N/m)  & 800   & 800   & 800 \\
   \qquad normal restitution coefficient        & 0.97  & 0.97  & 0.01 \\
   \qquad coefficient of friction               & 0.1   & 0.1   & 0.1 \\
   fluid properties & \\

   \qquad viscosity $\nu$~($\times 10^{-6}~\mathrm{m^2/}$)  
          & 1.0$\sim$5.4 & 1.0 & 1.0 \\
   \qquad density $\rho_f$~($\times 10^3~\mathrm{kg/m^3}$)  
          & 1.0 & 1.0 & 1.0 \\
   \hline
  \end{tabular}
 \end{center}
 \label{tab:param-all}
\end{table}

\begin{table}[!htbp]
  \caption{
  \label{tab:weight-fb}
  Breakdown of computational costs associated with different parts of fluidized bed simulations. For
  the both tests, the CPU times presented here are normalized by the time spent on the CFD
  part of the case using 256 processors.}
  \begin{center}
  \begin{tabular}{ccccc}
    \hline
    & $N_p = 4$ & 16 & 64 & 256 \\
    strong scaling  & & & & \\
    \hline
    CFD             &33.6(28\%)&7.7 (27\%)&1.9(23\%)&1.0(28\%)\\
    DEM             &89.0(71\%)&20.4(72\%)&5.9(74\%)&1.6(46\%)\\
    coupling        &3.1 (2\%) &0.4 (1\%) &0.2(3\%) &0.9(21\%)\\
    total           &125.7     &28.5      &8.0      &3.6 \\  
    \hline
    weak scaling    & & & & \\
    \hline
    CFD             &0.4(21\%)&0.3(27\%)&0.7(26\%)&1.0(31\%)\\
    DEM             &1.6(76\%)&1.7(72\%)&1.7(70\%)&1.9(56\%)\\
    coupling        &0.1(3\%) &0.0(1\%) &0.1(4\%) &0.4(13\%)\\
    total           &2.1      &2.1      &2.5      &3.3 \\  
    \hline
  \end{tabular}
  \end{center}
\end{table}

\begin{table}[!htbp]
  \caption{
  \label{tab:weight-st}
  Breakdown of computational costs associated with different parts of sediment transport
  simulations. For the both tests, the CPU times presented here are normalized by the time spent on
  the CFD part of the case using 8 processors.}
  \begin{center}
  \begin{tabular}{ccccc}
    \hline
    & $N_p = 8$ & 32 & 128 & 512\\
    strong scaling  & & & & \\
    \hline
    CFD         &74.4(47\%)&20.1(47\%)&4.4(45\%)&1.0(21\%)\\
    DEM         &35.8(23\%)&10.1(23\%)&2.8(29\%)&0.8(17\%)\\
    averaging   &45.7(29\%)&12.9(30\%)&2.5(25\%)&1.4(29\%)\\
    coupling    &1.2 (0\%) &0.2 (0\%) &0.1(1\%) &1.6(33\%)\\
    total       &157.0     &43.2      &8.0      &4.7 \\  
    \hline
    weak scaling    & & & & \\
    \hline
    CFD         &0.2(29\%)&0.3(37\%)&0.6(45\%)&1.0(48\%)\\
    DEM         &0.3(37\%)&0.3(36\%)&0.4(29\%)&0.4(20\%)\\
    averaging   &0.3(33\%)&0.2(27\%)&0.3(25\%)&0.4(20\%)\\
    coupling    &0.0(1\%) &0.0(1\%) &0.0(1\%) &0.2(11\%)\\
    total       &0.8      &0.9      &1.4      &2.1 \\  
    \hline
  \end{tabular}
  \end{center}

\end{table}

\end{document}